\documentclass[aps,prx,10pt,twocolumn,superscriptaddress]{revtex4-2}

\usepackage[english]{babel} 
\usepackage{amsmath} 
\usepackage{amssymb} 
\usepackage{graphicx} 
\usepackage{bm} 
\usepackage[hidelinks]{hyperref} 
\usepackage{mathtools} 
\usepackage{braket} 
\usepackage[capitalise]{cleveref} 
\usepackage{xcolor} 
\usepackage{xstring} 

\newcommand{\Norm}[1]{\left\lVert #1 \right\rVert}
\newcommand{\Abs}[1]{\left| #1 \right|}
\newcommand{\refRef}[1]{%
	\IfSubStr{#1}{,}{Refs.~\cite{#1}}{Ref.~\cite{#1}}%
}
\newcommand{\vvec}[1]{\bm{#1}}

\newcommand{\mmat}[1]{\text{\sffamily{#1}}}
\newcommand{\iim}{\mathrm{Im}}
\newcommand{\rre}{\mathrm{Re}}

\newcommand{\ii}{\mathrm{i}}

\newcommand{\dd}{\mathrm{d}}
\newcommand{\defEqual}{\overset{\text{def}}{=}}
\newcommand{\oneFtwo}[4]{%
{}_1F_2\!\left(
\genfrac{}{}{0pt}{}{#1}{#2,#3}
;\,#4
\right)
}
\newcommand{\twoFthree}[6]{%
{}_2F_3\!\left( 
\genfrac{}{}{0pt}{}{#1,#2}{#3,#4,#5}
;\,#6
\right)
}
\crefname{section}{Sec.}{Secs.}
\crefname{subsection}{Subsec.}{Subsecs.}
\crefname{equation}{Eq.}{Eqs.}
\crefname{figure}{Fig.}{Figs.}
\crefname{table}{Tab.}{Tabs.}
\crefname{appendix}{Appendix}{Appendices}
\crefname{chapter}{Chapter}{Chapters}

\hypersetup{
	pdfinfo={
		Author = {Baptiste Lorent, Jean-Marc Sparenberg and David Gaspard},
		Title = {Directionality emergence and localization in a quantum random Lorentz gas},
		Subject = {Scattering theory},
	},
	pdfborderstyle={/S/U/W 1},
	breaklinks=true
}%

\begin{document}%
\title{Directionality emergence and localization in a quantum random Lorentz gas}
\author{Baptiste Lorent}%
\email{baptiste.lorent@ulb.be}%
\author{Jean-Marc Sparenberg}%
\email{jmspar@ulb.ac.be}%
\affiliation{%
    Nuclear Physics and Quantum Physics, CP229, \\\href{https://ror.org/01r9htc13}{Université libre de Bruxelles (ULB)}, B-1050 Brussels, Belgium
}%
\author{David Gaspard}%
\email{david.gaspard@espci.fr}%
\affiliation{%
    \href{https://ror.org/00kr24y60}{Institut Langevin}, \href{https://ror.org/03zx86w41}{ESPCI Paris}, \href{https://ror.org/013cjyk83}{PSL University}, \href{https://ror.org/02feahw73}{CNRS}, F-75005 Paris, France
}%
\date{March 12, 2026}%

\begin{abstract}%
    \par The propagation of a spherical wave through a two-dimensional random Lorentz gas composed of small fixed scatterers is studied. Inspired by the Mott problem (how an initially isotropic quantum wave can give rise to a single particle-like track), we investigate, on a schematic model, whether such a directional behavior can emerge purely from the multiscattering process, without any explicit measurement or decoherence mechanism. Using the Foldy-Lax formalism, we derive the far-field angular behavior of the wavefunction, and introduce a directionality vector to quantify its anisotropy and identify its preferred direction. Numerical simulations reveal the existence of a strongly directional regime within a specific wavenumber range, which emerges from multiscattering with more than $100$ scatterers and which can be related to Anderson localization.
\end{abstract}%

\maketitle%

\section{Introduction}\label{sec:introduction}
\par Since the early days of quantum mechanics, the emergence of classical behavior in quantum systems has been the subject of continuous investigation. In this context, a particularly intriguing aspect is the apparent discrepancy between the microscopic wave description of radioactive processes proposed by Gamow in 1928 \cite{Gamow1928_Quantentheorie}, which models the emission of $\alpha$ radiation as a spherical wave emanating from the atomic nucleus (treated as a point source), and on the other hand, the observation of individual linear tracks in cloud chambers under the passage of such ionizing radiation, which indicates a particle-like nature. This example of a quantum-to-classical (or wave-to-particle) transition was originally studied by Mott in 1929 \cite{Mott1929_Wave} and is now known as the “Mott problem” \cite{Isham1981_Quantum,Wheeler1983_Quantum,Bell2004_Speakable,Cacciapuoti2007_Spin, Cacciapuoti2007_Solvable,Carlone2015_Model,Barletti2021_two-spin,Sparenberg2013_Decoherence,Sparenberg2018_Decoherence}. This problem can be divided into two parts: first, understanding why isotropic Gamow waves produce strongly directional tracks in gaseous detectors, and second, understanding why a single specific direction is selected among all a priori possible directions.

\par To address the first question, Mott introduced an \textit{ab initio} model, in the spirit of what he referred to as “unaided wave mechanics”. He included the degrees of freedom of the detector (in this case the excited states of the electrons of the gas atoms) in the complete wavefunction of the system, and showed that, among all possible excitation states, the probability of exciting two atoms simultaneously is maximal when they are aligned with the particle source. This can be understood as a consequence of the significant forward directionality of the differential cross-section of excitation of the particle on the atoms. This result can be generalized to a number $N$ of atoms composing the cloud chamber. Under a statistical average of the position and momentum of the gaseous particles surrounding the source, the system should predominantly populate excitation states where excited atoms are aligned with the source point. This approach thus explains the appearance of quasi-linear tracks left by $\alpha$-particles in cloud chambers. In this framework, the probability of macroscopic presence of these $\alpha$-particles is typically governed by a transport equation such as the Boltzmann transport equation \cite{Weinberg1958_Physical, Chandrasekhar1960_Radiative,Sigmund2006_Particle}. This approach is fully justified when the de Broglie wavelength of the quantum particle, $\lambda=2\pi/k$, is very small compared to the mean free path, $\ell$, because we can then assume independent scatterings \cite{Akkermans2007_Mesoscopic, Sheng2006_Introduction, Carminati2021_Principles}.

\par This transport equation describes very precisely not only the propagation of the particle in matter, but also its quantum {\em decoherence} \cite{Hornberger2003_Collisional, Vacchini2009_Quantum, Schlosshauer2019_Quantum, Diosi2023_Hybrid, Gaspard2022_master, Gaspard2023_Transverse,Pedalino2026_Probing}, that is the dynamical mechanism by which the quantum wave, via its multiple collisions with the gaseous environment, loses its quantum coherence and behaves classically, like a particle with a random direction \cite{Joos1984_Emergence,Teta2010_Classical}. From an experimental standpoint, several works have investigated the emergence of classical behavior in quantum systems, providing empirical evidence of decoherence \cite{Schlosshauer2019_Quantum,Hornberger2003_Collisional,Brune1996_Observing,Deleglise2008_Reconstruction,Pedalino2026_Probing}. Such experiments typically consist in probing quantum systems during an interaction, to observe the gradual disappearance of quantum features, or in externally manipulating the strength of those interactions, thus controlling the decoherence process. The latter approach is followed in \cite{Hornberger2003_Collisional}, where the authors study the emergence of collisional decoherence in the propagation of fullerene molecules through a gas. By gradually increasing the pressure of the gas, they tune the interaction strength, and measure the resulting interference vanishing, hence validating the theoretical approach of transport equations. Therefore, since the particle-detector interaction is short-range, the appearance of linear tracks is well understood.

\par The present paper is motivated by the second part of the Mott problem, namely the apparition of an apparently random {\em single} track during a {\em single} particle detection, which is much harder to solve. It raises the “which track” question and is a fundamental manifestation of the quantum measurement problem. A possible approach of this second problem, still in the spirit of the {\em unaided} quantum mechanics approach promoted by Mott, is to test whether the evolution of the spherical incident quantum wave on a single microscopic state of the detector could lead to a single detected linear track \cite{Sparenberg2013_Decoherence}. Of course, the above transport equation approach being only based on the local average density of scatterers, the microscopic positions of the atoms in the detector are not expected to play a discernible role in the probability of the $\alpha$-particle's presence, nor in its evolution over time. There are however special situations where the microscopic positions of the scatterers maintain their influence on the probability of the particle's presence. This is particularly the case when the wavelength is comparable to or greater than the mean free path. In this case, the assumption of independent scatterings no longer applies, and coherent effects dominate the incoherent propagation predicted by the usual transport equations. Disorder can then lead to the complete suppression of transport, a phenomenon known as Anderson localization \cite{Anderson1958_Absence,Lagendijk2009_Fifty,Skipetrov2014_Absence,Skipetrov2018_Ioffe-Regel}.

\par The objective of the present article is thus to study the influence of the microscopic positions of scatterers on the probability distribution of the quantum particle, and to check whether directionality emerges in some regimes, linked with Anderson localization or not. To do this, we depart from a realistic description of the Mott problem, which would require a quantum description of both the quantum particle and the gas on which it impinges, and which would restrict ourselves to the regime of incoherent and highly directional collisions. Here, we explore a different class of problems, where directionality and sensitivity to the microscopic state of the medium is expected to play a role.

\par For that, we consider a two-dimensional quantum random Lorentz gas model based on the Foldy-Lax formalism \cite{Foldy1945_Multiple,Lax1951_Multiple,Mishchenko2006_Multiple,Gaspard2022_Multiple,Joachain2023_Multiple}, in which the scatterers are randomly placed at fixed positions and modeled as points that scatter the incident wave isotropically and maximally. We assume that the gas is spherical and that the particle is emitted as a spherical wave from its center. The system is therefore spherically symmetric on average over the realizations of the disorder, making the whole model as spherical as possible. In order to highlight the influence of the positions of the scatterers, we introduce a new observable, directionality, which is the average direction taken by the quantum particle as it exits the gas. We will see that the particle exits the gas with a preferred direction much more often than would be expected classically. This directional preference is particularly marked when the limit of the Ioffe-Regel criterion is reached ($k \ell= 1$), indicating that this observable is very sensitive to localization effects.

\par The paper is organized as follows. In \cref{sec:angular_pdf_directionality}, we discuss some generalities about the angular probability density functions (PDFs), and we introduce the directionality observable, quantifying the anisotropy of these functions. \cref{sec:spherical_waves_scattering} is devoted to the analytical study of the scattering of a spherical wave, and to its application in the Foldy-Lax framework. \cref{sec:results} presents both analytical and numerical results for the propagation of spherical waves through two-dimensional gases. Finally, \cref{sec:conclusion} concludes this work and puts the results obtained for this simple model in a wider perspective.%

\section{Angular probability density functions and directionality}\label{sec:angular_pdf_directionality}
\par
\begin{figure}[b]%
    \includegraphics{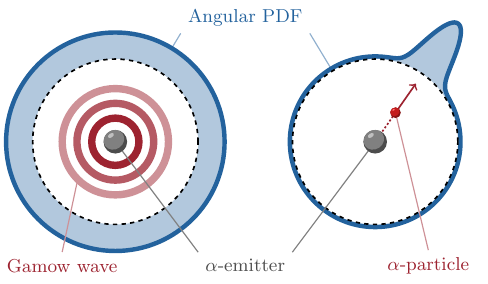}%
    \caption{Schematic comparison of the asymptotic behavior of a quantum spherical Gamow wave \textbf{(left)} and a classical $\alpha$-particle \textbf{(right)}. The very distinct nature of these two propagation modes is the motivation for the study of the angular probability density.}%
    \label{fig:particle_vs_wave}%
\end{figure}

 This first section aims at defining and characterizing an angular probability density function (also called angular PDF in this paper), as such objects naturally arise when studying the asymptotic behavior of a scattering state, especially in the context of the Mott problem \cite{Mott1929_Wave}. Indeed, as shown in \cref{fig:particle_vs_wave}, the asymptotic behavior of a spherical wave is isotropic, while a classical particle propagating in a fixed direction produces a sharply peaked distribution, ideally represented by an angular Dirac delta. The asymptotic angular PDF thus provides a natural ground for comparing a quantum spherical wave and a classical particle. The present section is especially devoted to the derivation of a quantity describing the directional behavior of these functions, effectively allowing to distinguish wave-like and particle-like propagation modes.

\par In arbitrary dimension $d$, an angular probability density function is a function of the form 
\begin{equation}\label{eq:angular_pdf_def}
    \frac{\dd P}{\dd \Omega} : \mathcal{S}^{d} \rightarrow \mathbb{R}^+, \vvec{\Omega} \mapsto \frac{\dd P}{\dd \Omega}(\vvec{\Omega}),
\end{equation}
that satisfies the normalization condition
\begin{equation}\label{eq:normalization_condition}
    \oint_{\mathcal{S}^{d}} \dd \Omega \, \frac{\dd P}{\dd \Omega}(\vvec{\Omega}) = 1,
\end{equation}
where $\mathcal{S}^{d}$ is the $d$-dimensional unit sphere, and $\vvec{\Omega}$ is the unit vector pointing towards the direction $\Omega$. The domain of $\dd P / \dd \Omega$ is the unit sphere $\mathcal{S}^{d}$, which is a non-euclidean space preventing the use of usual statistics tools such as the classical mean and variance. Instead, methods from directional statistics must be used \cite{Mardia2000_Directional,Fisher1987_Statistical,Fisher1993_Statistical,Pewsey2021_Recent}. Those tools are frequently used in the domain of protein structure analysis \cite{Hamelryck2006_Sampling,Ley2018_Applied,Boomsma2006_Graphical}, geophysical and oceanographic data analysis \cite{Mardia1981_Directional,Borradaile1997_Tectonic,Voronovich2017_Ocean}, and many others.

\par The first-order integral moment of \cref{eq:angular_pdf_def} is
\begin{equation}\label{eq:directionality_def}
    \vvec{w} \defEqual \oint_{\mathcal{S}^{d}} \dd \Omega \, \frac{\dd P}{\dd \Omega}(\vvec{\Omega}) \vvec{\Omega},
\end{equation}
and will be referred to as the \textit{directionality vector} in the following. It is given by the integral over $\mathcal{S}^{d}$ of the unit vectors $\vvec{\Omega}$ in each direction, weighted by the angular probability density function in that direction. Its orientation $\vvec{e}_w = \vvec{w} / ||\vvec{w}||$ corresponds to the mean direction towards which the angular probability density is directed. Correspondingly, its norm $w = ||\vvec{w}||$, that we call the \textit{directionality}, lies in the interval $[0,1]$, as
\begin{align}
    w \, & \ = \Norm{\oint_{\mathcal{S}_d} \dd \Omega \frac{\dd P}{\dd \Omega}(\vvec{\Omega}) \vvec{\Omega}}, \nonumber \\
    & \ \leq \oint_{\mathcal{S}_d} \dd\Omega \frac{\dd P}{\dd \Omega}(\vvec{\Omega}) ||\vvec{\Omega}||, \nonumber \\
    & \ \leq \oint_{\mathcal{S}_d} \dd\Omega \frac{\dd P}{\dd \Omega}(\vvec{\Omega}), \nonumber \\
    & \ \leq 1,
\end{align}
where we used respectively the triangular inequality, the fact that $\vvec{\Omega} $ has a unit norm, and \cref{eq:normalization_condition}. The more the anisotropy is pronounced towards a particular direction, the closer $w$ is to $1$. The limit case is the angular Dirac delta distribution $\dd P / \dd \Omega(\vvec{\Omega}) = \delta(\vvec{\Omega} - \vvec{\Omega_0})$, that has a directionality equal to $1$. On the contrary, an angular PDF that has little preferred direction will present a directionality close to $0$. The limit case is here the uniform distribution $\dd P / \dd \Omega(\vvec{\Omega}) = S_d^{-1}$, where $S_d$ is the surface area of the unit sphere $\mathcal{S}^{d}$. This distribution has a directionality equal to $0$. In this latter case, the directionality vector $\vvec{w}$ is null, and its orientation $\vvec{e}_w$ is undefined. It must be noted that, in addition to this trivial case, any distribution with no dipolar moment will also have a vanishing directionality vector.

\par The directionality $w$ is thus an indicator of the spread of the angular PDF, and is a good candidate to discriminate the two situations presented in \cref{fig:particle_vs_wave}. Indeed, the spherical wave has a directionality equal to $0$, while the classical particle has a directionality close or equal to $1$. However, the squared directionality $\vvec{w}^2$ is often preferable for analytical calculations, and will be used in the following instead of $w$, as it has the same bounds and interpretation as $w$. If the angular PDF is random due to the disorder of the scattering medium, then it is convenient to define the average squared directionality as
\begin{equation}\label{eq:directionality_squared}
    \langle \vvec{w}^2 \rangle  = \oint_{\mathcal{S}^{d}} \dd \Omega \oint_{\mathcal{S}^{d}} \dd \Omega' \, \left\langle \frac{\dd P}{\dd \Omega}(\vvec{\Omega}) \frac{\dd P}{\dd \Omega}(\vvec{\Omega}') \right\rangle \, \vvec{\Omega} \cdot \vvec{\Omega}',
\end{equation}
where $\langle \cdot \rangle$ denotes an averaging over the random configurations of the scattering medium. As shown in \cref{appsec:autocorrelation}, this expression is equivalent to
\begin{equation}\label{eq:directionality_squared_2}
    \langle \vvec{w}^2 \rangle  = \frac{S_{d-1}}{S_d} \int_0^\pi \dd\tau \, \sin^{d-2}(\tau) \cos(\tau) C(\tau).
\end{equation}
In the above expression, $C(\tau)$ is the angular correlation function which, in the case of a spherically symmetric system, is given by
\begin{equation}\label{eq:angular_autocorrelation_function_spherical_case}
    C(\tau) = S_d^2 \left\langle \frac{\dd P}{\dd \Omega}(\vvec{\Omega}) \frac{\dd P}{\dd \Omega}(\vvec{\Omega}') \right\rangle
\end{equation}
where $\vvec{\Omega}$ and $\vvec{\Omega}'$ are two arbitrary directions separated by the angle $\tau$.

\par In the following, we will use these quantities to quantify the anisotropy of the angular PDFs that will be encountered in the study of the scattering of spherical waves in a Lorentz gas. Our main goal is to assess whether an initially spherical wave with $w=0$ can evolve into a directional state with $w \approx 1$, as this would indicate a particle-like behavior emergence.%

\section{Scattering of spherical waves in arbitrary dimension}\label{sec:spherical_waves_scattering}
\par As mentioned in the introduction, this work's specificity is to analyze the scattering of a spherical wave off a potential. The standard theory established for incident plane waves, as well as most of the quantities used in it, must then be revised. This section therefore aims at providing a meaningful mathematical formalism for this case. In particular, we introduce an important quantity, the \textit{spherical scattering amplitude} (SSA), which plays a central role in the following. \cref{subsec:spherical_formalism} gives an overview of the formalism of spherical waves scattering, while \cref{subsec:foldy_lax} introduces the Foldy-Lax model. Throughout this section, $d$ represents the arbitrary dimension of the space, and we use the reduced unit system $\hbar^2 / 2m = 1$.

\subsection{Formalism of spherical waves scattering}\label{subsec:spherical_formalism}
\par The beginning of this Subsection follows \refRef{Gaspard2022_Multiple} by defining the Green functions associated with the problem, and their main properties. Further definitions and proofs can be found in that reference.

\subsubsection{Free Green function}
\par The free Green function associated with the time-independent Schrödinger equation describes the propagation of a quantum wave emitted from a point source in absence of any potential, and is defined as the solution of
\begin{equation}
    \left[ \vvec{\nabla}^2_{\vvec{r}} + k^2 \right] G(k, \vvec{r} | \vvec{x}_0) = \delta^{(d)} (\vvec{r} - \vvec{x}_0),
\end{equation}
where $k$ is the wavenumber, and $\vvec{x}_0$ is the position of the source point. This function can be extended to values of $k \in \mathbb{C}$,  which allows to define the retarded and advanced free Green functions
\begin{align}
    G^{\pm}(k, \vvec{r} | \vvec{x}_0) \ & = \lim_{\varepsilon \rightarrow 0^+} \bra{\vvec{r}} \frac{1}{k^2 \pm \ii \varepsilon - \hat{\vvec{p}}^2} \ket{\vvec{x}_0}, \nonumber \\
    & = \pm\frac{1}{4 \ii} \left(\frac{k}{2 \pi r}\right)^{\frac{d-2}{2}} H^{\pm}_{\frac{d-2}{2}}(k r), 
\end{align}
where $\hat{\vvec{p}} = -\ii \vvec{\nabla}_{\vvec{r}}$ is the momentum operator, related to the wavenumber $k$, $r = ||\vvec{r} - \vvec{x}_0||$ is the distance between the point $\vvec{r}$ and the source point $\vvec{x}_0$, and $H^{\pm}_{\nu}$ are the outgoing ($+$) and incoming ($-$) Hankel functions. We finally define the free local density of states
\begin{align}\label{eq:free_local_density_of_states}
    \mu_0(k) \ & = \bra{\vvec{r}} \delta^{(d)} (k^2 - \hat{\vvec{p}}^2) \ket{\vvec{r}}, \nonumber \\
    & = -\frac{1}{\pi} \iim [G^{+}(k, \vvec{r} | \vvec{r})], \nonumber\\
    & = \frac{S_d k^{d-2}}{2 (2 \pi)^d}.
\end{align}
It is worth noting that the free outgoing Green function asymptotically satisfies the relations
\begin{align}
    G^+(k, \vvec{r} | \vvec{x}) \ & \xrightarrow[]{r \rightarrow \infty} G^+(k, \vvec{r} | \vvec{x}_0) e^{- \ii k \vvec{\Omega} \cdot (\vvec{x} - \vvec{x}_0)}, \label{eq:free_green_asymptotic_1} \\
    \Abs{G^+(k, \vvec{r} | \vvec{x}_0)}^2 \ & \xrightarrow[]{r \rightarrow \infty} \frac{\pi \mu_0(k)}{k S_d r^{d-1}}. \label{eq:free_green_asymptotic_2}
\end{align}

\subsubsection{Full Green function}
\par On the other hand, the full Green function associated with the Schrödinger equation is the wavefunction that results from the scattering process of the wave emitted from a point source and propagating through a potential $U(\vvec{r})$. It satisfies the equation
\begin{equation}
    \left[ \nabla^2_{\vvec{r}} + k^2 - U(\vvec{r}) \right] \psi(k, \vvec{r} | \vvec{x}_0) = \delta^{(d)} (\vvec{r} - \vvec{x}_0).
    \label{eq:schrodinger}
\end{equation}
Again, by extending the wavenumber domain to complex numbers, one defines
\begin{equation}
    \psi^{\pm}(k, \vvec{r} | \vvec{x}_0) = \lim_{\varepsilon \rightarrow 0^+} \bra{\vvec{r}} \frac{1}{k^2 \pm \ii \varepsilon - \hat{\vvec{p}}^2 - U(\hat{\vvec{r}})} \ket{\vvec{x}_0}.
\end{equation}
Analogously to what is done for the free Green function, we define the full local density of states
\begin{align}
    \mu(k, \vvec{r}) \ & = \bra{\vvec{r}} \delta^{(d)} (k^2 - \hat{\vvec{p}}^2 - U(\hat{\vvec{r}})) \ket{\vvec{r}}, \nonumber \\
    & = -\frac{1}{\pi} \iim [\psi^{+}(k, \vvec{r} | \vvec{r})].
    \label{eq:full_local_density_of_states}
\end{align}
Next sections are devoted to derive the full Green function $\psi^+(k, \vvec{r} | \vvec{x}_0)$ (which will also be referred to as the wavefunction in the following), and to characterize its asymptotic behavior, in order to study the directionality that emerges from the scattering process.

\subsubsection{Spherical scattering amplitude}
\par To describe the behavior of the full Green function far from the emitter $\vvec{x}_0$, we introduce the SSA $A(\vvec{\Omega})$, as the angular asymptotic behavior of the wavefunction
\begin{equation}\label{eq:scattering_amplitude_definition}
    \psi^+(k, \vvec{r} | \vvec{x}_0) \xrightarrow[]{r \rightarrow \infty} A(\vvec{\Omega}) G^+(k, \vvec{r} | \vvec{x}_0),
\end{equation}
where $\vvec{\Omega} = (\vvec{r}-\vvec{x}_0)/\Norm{\vvec{r}-\vvec{x}_0}$ is the direction. The SSA can be interpreted as the weighting factor by which one should multiply the incident spherical wave $G^+(k, \vvec{r} | \vvec{x}_0)$, which does not depend on $\vvec{\Omega}$, to get the full Green function $\psi^+(k, \vvec{r} | \vvec{x}_0)$, when the distance from $\vvec{x}_0$ increases to infinity.

\par A constraint on $A(\vvec{\Omega})$ can be derived from the conservation of the quantum probability flux given by $\vvec{J}(\vvec{r}) = \iim[\psi^*(\vvec{r}) \vvec{\nabla}_{\vvec{r}} \psi(\vvec{r})]$. Let $\mathcal{B}_0$ be a ball of radius tending to zero, and $\mathcal{B}_\infty$ a ball of radius tending to infinity, both centered around $\vvec{x}_0$. As the only source is the emitter located at $\vvec{x}_0$, the probability current $I_0$ crossing the surface $\partial \mathcal{B}_0$ must be equal to the probability current $I_\infty$ crossing the surface  $\partial \mathcal{B}_\infty$. In other words, 
\begin{equation}\label{eq:conservation_integral}
     \underbrace{\oint_{\partial \mathcal{B}_\infty} \dd \vvec{S} \cdot \vvec{J}(\vvec{r})}_{I_\infty} = \underbrace{\oint_{\partial \mathcal{B}_0} \dd \vvec{S} \cdot \vvec{J}(\vvec{r})}_{I_0},
\end{equation}
where $\dd \vvec{S} = \vvec{\Omega} \, r^{d-1}\, \dd \Omega$ is a spherical surface element. The two integrals present in \cref{eq:conservation_integral} can be rewritten as \cref{eq:I_infinity,eq:I_zero} respectively (see \cref{appsec:probability_currents}). Inserting these new expressions in the equality gives
\begin{equation}\label{eq:normalization_C}
    \oint_{\mathcal{S}_d} \dd \Omega \, |A(\vvec{\Omega})|^2 = S_d \frac{\mu(k, \vvec{x}_0)}{\mu_0(k)}.
\end{equation}
This condition ensures that the SSA is a square integrable function, and thus that a unique angular probability density function can be associated to the far-field wavefunction. In order to fulfill the normalization condition of \cref{eq:normalization_condition}, such a quantity must be defined as
\begin{equation}\label{eq:asymptotic_directional_proba_distribution_def}
    \frac{\dd P}{\dd \Omega}(\vvec{\Omega}) = \frac{1}{S_d} \frac{\mu_0(k)}{\mu(k, \vvec{x}_0)} \Abs{A(\vvec{\Omega})}^2.
\end{equation}
This quantity is the angular PDF for the detection of the scattered wave at an angle $\vvec{\Omega}$ far from the gas. Indeed, it is proportional to $\Abs{A(\vvec{\Omega})}^2$, which represents the asymptotic angular part of the full Green function at large distance. Therefore, for a spherical detector placed on the sphere $\partial \mathcal{B}_\infty$ (infinitely far away from $\vvec{x}_0$), the probability of detecting the scattered wave in the direction $\vvec{\Omega}$ is given by $\dd P/\dd \Omega(\vvec{\Omega})$. In absence of any potential around the source, the scattering state is the free Green function $G^+(k, \vvec{r} | \vvec{x}_0)$, which does not depend on $\vvec{\Omega}$, and the angular PDF is thus the uniform density $1/S_d$. Any deviation from this uniform density is a result of the scattering process. In the following, we will especially focus on the directionality that emerges in the angular PDF from the scattering process.

\subsection{Application to the Lorentz gas model}\label{subsec:foldy_lax}
\par The Foldy-Lax formalism \cite{Foldy1945_Multiple,Lax1951_Multiple,Gaspard2022_Multiple} allows to compute the wavefunction of the scattered wave for a propagation in a Lorentz gas. This medium is described by a potential of the shape
\begin{equation}
    U(\vvec{r}) = \sum_{i=1}^{N} u(\vvec{r} - \vvec{x}_i), 
\end{equation}
where $N$ denotes the number of scatterers in the gas, $\vvec{x}_1, \vvec{x}_2, ..., \vvec{x}_N$ are their positions, and $u(\vvec{r})$ is the potential associated to each scatterer. Its precise shape has no importance, but it must be a short-range potential as compared with the wavelength, meaning that $u(\vvec{r})$ must vanish beyond a certain cutoff radius $b$ such that $kb \ll 1$. This condition ensures that each individual collision only involves $s$-waves. In this model, the wavefunction satisfying \cref{eq:schrodinger} is
\begin{align}\label{eq:foldy_lax_wave_fct_1}
    \psi^+(k, \vvec{r} | \vvec{x}_0) \ & = G^+(k, \vvec{r} | \vvec{x}_0) + \sum_{j=1}^{N} a_j G^+(k, \vvec{r} | \vvec{x}_j).
\end{align}
The full wavefunction $\psi^+(k, \vvec{r} | \vvec{x}_0)$ is the sum of the incident spherical Gamow wave originating from $\vvec{x}_0$, and the $N$ spherical waves originating from the gaseous scatterers at $\vvec{x}_j \ (j \neq 0)$, each modulated by a local scattering amplitude $a_j$. The self-consistent equation for the amplitudes $a_j$ reads \cite{Gaspard2022_Multiple}
\begin{equation}\label{eq:foldy_lax_linear_system}
    \mmat{M}
    \begin{pmatrix}
        a_1\\
        \vdots\\
        a_N
    \end{pmatrix}
    = 
    \begin{pmatrix}
        G^+(k, \vvec{x}_1 | \vvec{x}_0)\\
        \vdots\\
        G^+(k, \vvec{x}_N | \vvec{x}_0)
    \end{pmatrix},
\end{equation}
where the elements of the multiple-scattering matrix $\mmat{M}$ are defined as
\begin{equation}\label{eq:foldy_lax_multiple_scattering_matrix}
    M_{ij} \defEqual F(k)^{-1}\delta_{ij} - G^+(k, \vvec{x}_i | \vvec{x}_j) (1 - \delta_{ij}).
\end{equation}
In this equation, $F(k)$ is the point scattering amplitude. It depends on the scattering model used to describe the collision, and is isotropic due to the point nature of the scatterers. Using the far-field behavior of \cref{eq:free_green_asymptotic_1} in \cref{eq:foldy_lax_wave_fct_1}, and comparing with \cref{eq:scattering_amplitude_definition}, one finds that the SSA in the Foldy-Lax model satisfies
\begin{equation}\label{eq:foldy_lax_scattering_amplitude}
    A(\vvec{\Omega}) = 1 + \sum_{j=1}^N A_j(\vvec{\Omega}),
\end{equation}
where we defined $A_j(\vvec{\Omega}) = a_j e^{-\ii  k \vvec{\Omega} \cdot (\vvec{x}_j - \vvec{x}_0)}$ for $j = 1, \ldots, N$. Therefore, following \cref{eq:asymptotic_directional_proba_distribution_def}, the detection probability density function is, in the Foldy-Lax model, given by
\begin{align}\label{eq:foldy_lax_pdf}
    \frac{\dd P}{\dd \Omega}(\vvec{\Omega}) &= \frac{1}{S_d} \frac{\mu_0(k)}{\mu(k, \vvec{x}_0)} \left[ 1 + 2 \sum_{j=1}^N \rre [A_j(\vvec{\Omega})] \right. \nonumber \\
    & \left. + \sum_{j,l=1}^N A_j(\vvec{\Omega}) A_l^*(\vvec{\Omega}) \right].
\end{align}
In this expression the first term is due to the incident spherical wave only, while the second term accounts for the interference between the incident wave and each scattered wave. The last term describes the interference between each pair of scattered waves. The factor $\mu(k, \vvec{x}_0)$ that appears in the right-hand side of this equation can be explicitly expressed in the Foldy-Lax framework by injecting \cref{eq:foldy_lax_wave_fct_1} in \cref{eq:full_local_density_of_states}. It reads
\begin{equation}\label{eq:foldy_lax_mu_ratio}
    \mu(k, \vvec{x}_0) = \mu_0(k) - \frac{1}{\pi}\iim \sum_{j=1}^{N} a_j G^+(k, \vvec{x}_0 | \vvec{x}_j).
\end{equation}
The directional behavior of the angular PDF can be quantified by the directionality vector $\vvec{w}$ defined in \cref{eq:directionality_def}. The full computation of $\vvec{w}$ is provided in \cref{appsec:analytical_w}, and its final analytical expression is
\begin{align}\label{eq:analytical_foldy_lax_w_final_form}
    \vvec{w} =\hspace{4pt}& \frac{2k}{d} \frac{\mu_0(k)}{\mu(k, \vvec{x}_0)} \left[ \sum_{\substack{l=1\\\phantom{}}}^N \iim (a_l) (\vvec{x}_l - \vvec{x}_0) Z_d(k \Norm{\vvec{x}_l - \vvec{x}_0}) \right. \nonumber \\
    & \left. + \sum_{\substack{j=1\\l>j}}^N \iim (a_l a_j^*) (\vvec{x}_l - \vvec{x}_j) Z_d(k \Norm{\vvec{x}_l - \vvec{x}_j}) \right],
\end{align}
where the function $Z_d(z)$ is defined in \cref{eq:integral_Z}. The vector $\vvec{w}$ is thus the summation of $N$ single-scatterer contributions (first summation), and of $N(N-1)/2$ two-scatterers contributions (second summation). This analytical expression can be useful for numerical simulations, in order to avoid performing a spherical integration over the angular PDF to compute $\vvec{w}$. However, its complexity prevents from easily deriving further statistical quantities such as the disorder-averaged squared directionality $\langle \vvec{w}^2 \rangle$.%

\section{Study of the two-dimensional propagation}\label{sec:results}
\par The present section is devoted to the analytical and numerical study of the directionality of a wavefunction propagating through a two-dimensional Lorentz gas. In such a configuration, the free outgoing Green function reads
\begin{equation}
    G^+(k, \vvec{r} | \vvec{x}_0) = \frac{1}{4 \ii } H^{+}_{0}(k r),
\end{equation}
where $r = ||\vvec{r} - \vvec{x}_0||$ is the distance between $\vvec{r}$ and the source point $\vvec{x}_0$, and $k$ is the wavenumber. Moreover, to align with the conventions commonly used in two-dimensional systems, the angular coordinate will be denoted by $\theta$ instead of $\Omega$. The unit vector on the circle is $\vvec{\Omega} = (\cos \theta, \sin \theta)$. As it only depends on the angle $\theta$, the quantities that depend on $\vvec{\Omega}$ will simply be expressed in terms of $\theta$ instead. 

\par Throughout this section, we use the parameter-free scattering model introduced in \refRef{Gaspard2022_Multiple,Gaspard2024_Effective}, designed to maximize the total cross-section. In this model, the single-collision scattering amplitude and the corresponding total cross-section are
\begin{equation}\label{eq:single_scattering_amplitude_and_cross_section}
    F(k) = -4\ii, \quad \sigma_{\text{pt}}(k) = \frac{4}{k}.
\end{equation}

\par This section is divided into four main parts. An analytical study of the single-scatterer system is first presented in \cref{subsec:single_particle}. Then, the many-scatterer system is considered in \cref{subsec:multiple_particles}, where the directionality is studied through numerical simulations and statistical analysis. The particular case of the ballistic regime is analytically discussed in \cref{subsec:ballistic_regime}, while a rudimentary analytical model of the directionality for the multiple-scatterer case is finally proposed in \cref{subsec:random_model}.

\subsection{Analytical study of the single-scatterer system}\label{subsec:single_particle}
\par The scattering through a single-scatterer system is a particularly simple case, for which an analytical solution can be derived. Moreover, such an analysis allows one to build an intuition about the wavefunction behavior in a simple setting, as well as to illustrate the action of the newly introduced vector $\vvec{w}$, which plays a central role in this work. The situation consists in a spherical wave emitted from $\vvec{x}_0$, and isotropically scattered by a single scatterer located at $\vvec{x}_1$. Due to the spherical symmetry of the problem, the norm of the directionality vector $\vvec{w}$ does not depend on the direction of $\vvec{x}_1 - \vvec{x}_0$, but only on the distance $s = \Norm{\vvec{x}_1 - \vvec{x}_0}$ between the emitter and the scatterer. We thus choose to place the unique scatterer at $\vvec{x}_1 = \vvec{x}_0 + s \vvec{e}_x$. The unique scattering parameter that will influence the directionality of the resulting wave is the product $ks$.

\par
\begin{figure*}[!]%
    \includegraphics{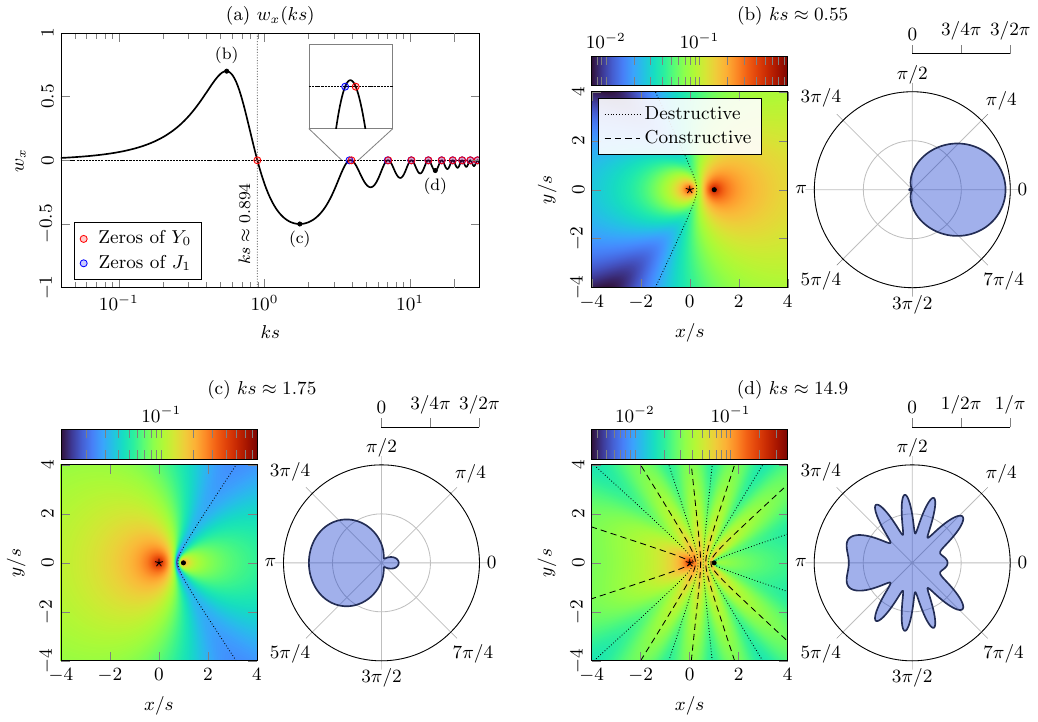}%
    \caption{(a) Evolution of the directionality $w_x$ in a single-scatterer gas as a function of $ks$. The red circles are the zeros of $Y_0$ and the blue circles are the zeros of $J_1$. (b-d) Square modulus of the full wavefunction $|\psi(k, \vvec{r} | \vvec{x}_0)|^2$ and angular PDF $\dd P (\theta) / \dd \theta$, respectively for (b) $ks \approx 0.55$ (focalizer regime), (c) $ks \approx 1.75$ and (d) $ks \approx 14.9$ (reflector regimes). The spherical wave source is represented by a star, and the unique scatterer by a black dot. The dotted and dashed lines represent the hyperbolic fringes satisfying \cref{eq:single_particle_hyperbolic_fringes}.}%
    \label{fig:single_particle_directionality}%
\end{figure*}

 The analytical expression of \cref{eq:analytical_foldy_lax_w_final_form} can be simplified in this case, as the only non-zero contribution to the summation corresponds to the pair $j=0$ and $l=1$. Moreover, the Foldy-Lax system \cref{eq:foldy_lax_linear_system} boils down to a single equation, whose solution is $a_1 = -4 \ii  G^+(k, \vvec{x}_1 | \vvec{x}_0)$. It should be noted that there is no Foldy-Lax amplitude associated with $\vvec{x}_0$, because there is no scatterer present at that location. Injecting this in the analytical expression of the directionality vector, one finds, upon simplification,
\begin{equation}\label{eq:single_particle_directionality}
    \vvec{w} = \frac{-2 Y_0(ks) J_1(ks)}{1 + Y_0^2(ks) - J_0^2(ks)} \vvec{e}_x,
\end{equation}
where $J_\nu$ and $Y_\nu$ are the Bessel functions of the first and second kind, respectively. As expected from the symmetry of the problem, this vector is oriented along the $x$-axis. The $x$ component of this vector, written $w_x$, provides important information about the scattering process. Indeed, if $w_x$ is positive, this means that the scatterer acts like a focalizer, as the wave is preferentially scattered towards its direction. On the other hand, a negative value of $w_x$ indicates that the wave is rather scattered in the opposite direction, and thus that the scatterer acts like a reflector. The behavior of $w_x$ is illustrated in \cref{fig:single_particle_directionality}(a) as a function of the product $ks$. It is interesting to categorize the zeros of this function by their origin, whether they are due to $Y_0$ (in red on the figure), or to $J_1$ (in blue). The generalized version of the interlacing theorem \cite{DLMF, Watson1996_Treatise} developed in \refRef{Palmai2011_Interlacing} ensures that the zeros of $Y_0$ and $J_1$ are interlaced, meaning that they strictly alternate as $ks$ increases on the positive real axis, which is visible on the figure. Moreover, because these two Bessel functions have the same asymptotic development, their respective zeros get asymptotically close to each other. More precisely, the $n$-th zero of $J_1$ gets closer and closer to (but is always smaller than) the $(n+1)$-th zero of $Y_0$, as $ks$ increases. This is also visible in \cref{fig:single_particle_directionality}(a), and means that, except for the first one that is due to $Y_0$, the zeros of $w_x(ks)$ occur in pairs.

\par Before the first positive zero of $w_x(ks)$, which is located at $ks \approx 0.894$, the value of $w_x$ is positive, meaning that the wave is preferentially scattered towards the direction of the scatterer (focalizer regime). After this first zero, the value of $w_x$ is negative, and the wave is preferentially scattered towards the opposite direction (reflector regime). From this point onwards, the function remains mainly negative, and only becomes positive in the very small intervals between pairs of successive roots, as can be seen on the zoomed panel of \cref{fig:single_particle_directionality}(a) (more precisely, between the $n$-th zero of $J_1$ and the $(n+1)$-th zero of $Y_0$). This means that, for $ks > 0.894$, the scatterer tends to reflect the wave in the opposite direction for almost all values of $ks$.

\par This behavior can be qualitatively understood by analyzing the full wavefunction $\psi^+(k, \vvec{r} | \vvec{x}_0)$, which results from the superposition of the incident and scattered spherical waves. Under the conditions
\begin{equation}\label{eq:fringes_hypothesis}
    k \Norm{\vvec{r} - \vvec{x}_0}, k \Norm{\vvec{r} - \vvec{x}_1}, ks \gg 1,
\end{equation}
the asymptotic form of the Hankel function \cite{Watson1996_Treatise,DLMF} allows one to express this wavefunction as
\begin{align}\label{eq:single_particle_wave_function_asymptotic}
    \psi^+(k, \vvec{r} | \vvec{x}_0) & \approx \alpha \exp[\ii(k \Norm{\vvec{r} - \vvec{x}_0} - 3\pi/4)] \nonumber \\
    & + \beta \exp[\ii(ks - k \Norm{\vvec{r} - \vvec{x}_1})],
\end{align}
where $\alpha$ and $\beta$ are positive coefficients. Under these conditions, the constructive and destructive interferences occur when the phase difference between the two terms of \cref{eq:single_particle_wave_function_asymptotic} is an even or odd multiple of $\pi$, respectively. Upon simplification, these conditions boil down to
\begin{equation}\label{eq:single_particle_hyperbolic_fringes}
    \frac{\Norm{\vvec{r} - \vvec{x}_0} - \Norm{\vvec{r} - \vvec{x}_1}}{s} = \frac{1}{ks} \left( \frac{3\pi}{4} - n\pi \right) + 1
\end{equation}
where $n$ is even for constructive fringes and odd for destructive ones. These equations describe branches of hyperbolae with foci at $\vvec{x}_0$ and $\vvec{x}_1$, and are solvable only when their right-hand side lies within $[-1, 1]$, thus limiting the observable fringe orders $n$ to the set
\begin{equation}\label{eq:fringe_order}
    \mathcal{N} = \left\{\, n \in \mathbb{Z} \;\middle|\; 
    1 \leq n \leq \frac{3}{4} + \frac{2ks}{\pi} \,\right\}.
\end{equation}
When $ks < \pi/8$, there is no interference fringe, and the wavefunction remains relatively isotropic, hence explaining the low value of $w_x$ at small $ks$. Then, the first destructive fringe ($
n=1$) appears when $ks = \pi/8$, initially in the direction opposite to the scatterer. The wavefunction is therefore suppressed in this direction and enhanced in the forward direction, and the angular PDF is thus dominated by an important lobe in the forward direction, which explains the positive value of $w_x$ in this regime (focalizer regime). As $ks$ further increases, this destructive fringe progressively opens towards the scatterer's direction. During this drift, the directionality becomes maximal at $ks \approx 0.55$. The wavefunction in this situation is shown in \cref{fig:single_particle_directionality}(b), where the forward lobe is clearly visible. The hyperbola deduced from \cref{eq:single_particle_hyperbolic_fringes} does not perfectly match the real destructive fringe of the wavefunction visible on the figure, because $ks$ is not large enough to satisfy \cref{eq:fringes_hypothesis}, but the qualitative behavior is well captured. Then, when $ks$ approaches $\pi/4$, the hyperbola becomes approximately vertical, corresponding to a nearly symmetric configuration where forward and backward scattering balance each other. In the exact solution, however, the directional zero occurs slightly later, at $ks \approx 0.894$, again because $ks$ is not large enough to satisfy \cref{eq:fringes_hypothesis}. \cref{fig:single_particle_directionality}(c) shows the situation for $ks \approx 1.75$. The destructive fringe is now oriented in the scatterer's direction, and the forward lobe is gradually replaced by a dominant backward lobe. In this regime, the directionality is therefore negative (reflector regime). Then, when $ks = 5\pi/8$, the second constructive fringe ($n=2$) appears in the backward direction, and gradually opens up for increasing values of $ks$. As $ks$ further increases, new fringes emerge one after the other, leading to alternating lobes and valleys in the backward direction. This is shown in \cref{fig:single_particle_directionality}(d), where the wavefunction is shown for $ks \approx 14.9$. The destructive fringe is now located on the right, and tends to diminish the wavefunction in that direction, explaining the negative value of $w_x$ in that region. The oscillation of $w_x$ as $ks$ increases is thus a direct consequence of the appearance of new fringes, which alternate between constructive and destructive interference.

\subsection{Numerical study of many-scatterer gas}\label{subsec:multiple_particles}
\par A Lorentz gas is a system composed of a large number of obstacles which scatter the incident wave ($N \gg 1$). Contrary to the single-scatterer system, an analytical solution cannot be derived for the general case. Therefore, numerical simulations are required to study the directionality of the wavefunction. The physical situation considered here consists in a disk-shaped gas of radius $R$, centered on the origin $\vvec{0}$, and containing $N$ point-like scatterers located at random positions $\vvec{x}_1, \vvec{x}_2, ..., \vvec{x}_N$. The mean interscatterer distance $\varsigma$ is related to the radius of the gas via the relation
\begin{equation}\label{eq:detector_radius}
    R = \varsigma \sqrt{\frac{N}{\pi}},
\end{equation}
leading to a scatterer density $n = \varsigma^{-2}$ for any value of $N$. With the single-scattering cross-section chosen in \cref{eq:single_scattering_amplitude_and_cross_section}, the mean free path satisfies the relation \cite{Gaspard2022_Multiple}
\begin{equation}\label{eq:diffusion_length}
    \frac{\ell}{\varsigma} = \frac{1}{n\varsigma\sigma_{\text{pt}}(k)} = \frac{k \varsigma}{4}.
\end{equation}

\par The two parameters that characterize the macroscopic state are thus $N$ and $k\varsigma$. For each couple of macroscopic parameters $(N, k\varsigma)$, a large number of microscopic configurations were simulated. Each simulation consists in randomly generating a set of $N$ points uniformly distributed in the disk of radius $R$, computing the solution of the $N \times N$ system of \cref{eq:foldy_lax_linear_system} for the coefficients $a_j$, and then subsequently evaluating the directionality vector $\vvec{w}$. As discussed in \cref{appsec:fourier_analysis}, increasing the wavenumber leads to a more oscillatory behavior of $\dd P(\theta) / \dd \theta$. In the low-wavenumber regime ($k \varsigma < 20$), where the oscillations remain moderate, $\vvec{w}$ is computed through a direct numerical evaluation of \cref{eq:directionality_def}. By contrast, in the high-wavenumber regime ($k \varsigma > 20$), the rapid oscillations of the PDF makes numerical integration inefficient and inaccurate, and the analytical expression \cref{eq:analytical_foldy_lax_w_final_form} is used instead.

\par
\begin{figure}[!t]%
    \hspace{-.25cm}\includegraphics{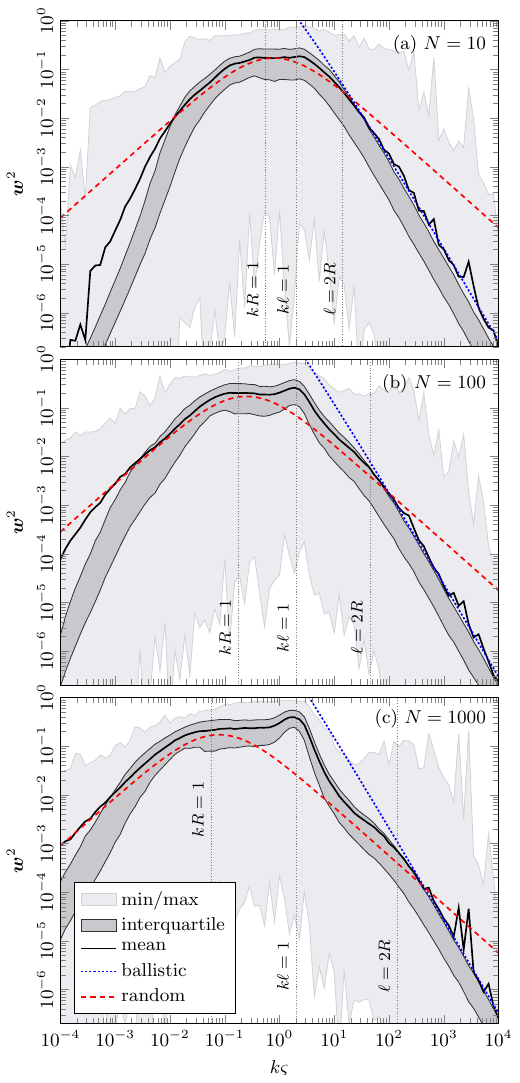}%
    \caption{Evolution of $\vvec{w}^2$ as a function of $k \varsigma$, for a gas containing $10$, $100$, and $1000$ scatterers (panels (a), (b) and (c) respectively). For each value of $k \varsigma$, $5000$ random configurations were simulated. The average value of $\vvec{w}^2$, the interquartile and the full range of the distribution are shown. The blue dotted curves represent the analytical prediction in the ballistic regime from \cref{eq:ballistic_regime_w}, while the red dashed curves show the corresponding prediction for $\langle \vvec{w}^2 \rangle$ associated with the random model of \cref{eq:random_model_w}. The vertical line at $k \ell = 1$ indicates the position of the Ioffe-Regel threshold, around which the strong directionality peak emerges.}%
    \label{fig:w_vs_ksigma}%
\end{figure}

 Before analyzing the results of these simulations, we first note that the problem is spherically symmetric in the sense of \cref{eq:isotropic_average_pdf}. Indeed, averaging over all realizations of the gas is equivalent to averaging over all rotations of a given configuration around the source point. As a result, the angular PDF is isotropic on average, $\dd P(\theta) / \dd \theta = 1/2 \pi$. This implies that the directionality vector is zero on average, $\langle \vvec{w} \rangle = \vvec{0}$. However, the expected value of its squared norm $\langle \vvec{w}^2 \rangle$ is typically nonzero. The evolution of $\langle \vvec{w}^2 \rangle$ with $k \varsigma$ is presented in \cref{fig:w_vs_ksigma} for $N = 10$, $100$ and $1000$ scatterers. The figure also shows the interquartile and full range of the distribution for each value of $N$ and $k\varsigma$. For all values of $N$, the graph of $\langle \vvec{w}^2 \rangle$ can be separated in two distinct contributions. The first one, which constitutes the background, is a bell-shaped curve which translates towards $k = 0$ as $N$ grows, without changing the height of its plateau $\langle \vvec{w}^2 \rangle \approx 0.2$. The second contribution is an additional directionality peak, located around $k \varsigma \approx 2$, whose maximum value significantly increases as $N$ increases. In virtue of \cref{eq:diffusion_length}, the peak position coincides with the localization threshold
\begin{equation}
    k\ell \cong 1, 
\end{equation}
corresponding to the Ioffe-Regel criterion \cite{Sheng2006_Introduction,Skipetrov2018_Ioffe-Regel,Ahn2022_Anderson}. This suggests that this peak is a manifestation of localization due to multiple scattering of the wave. In addition, the peak position also coincides with the regime where the propagation is largely dominated by backscattering (see \cref{fig:single_particle_directionality}(c)), a phenomenon known to give rise to localization effects \cite{Akkermans2007_Mesoscopic}. \cref{fig:w_vs_ksigma}(a) shows that a gas with $10$ scatterers is largely dominated by the background contribution, as no peak is visible around the Ioffe-Regel threshold. In \cref{fig:w_vs_ksigma}(b-c), we observe that the larger the gas, the higher the directionality peak. This means that, for $k\ell \cong 1$, a random configuration of the gas is, in average, more and more likely to produce a directional wave as $N$ increases.

\begin{figure*}[p]%
    \includegraphics{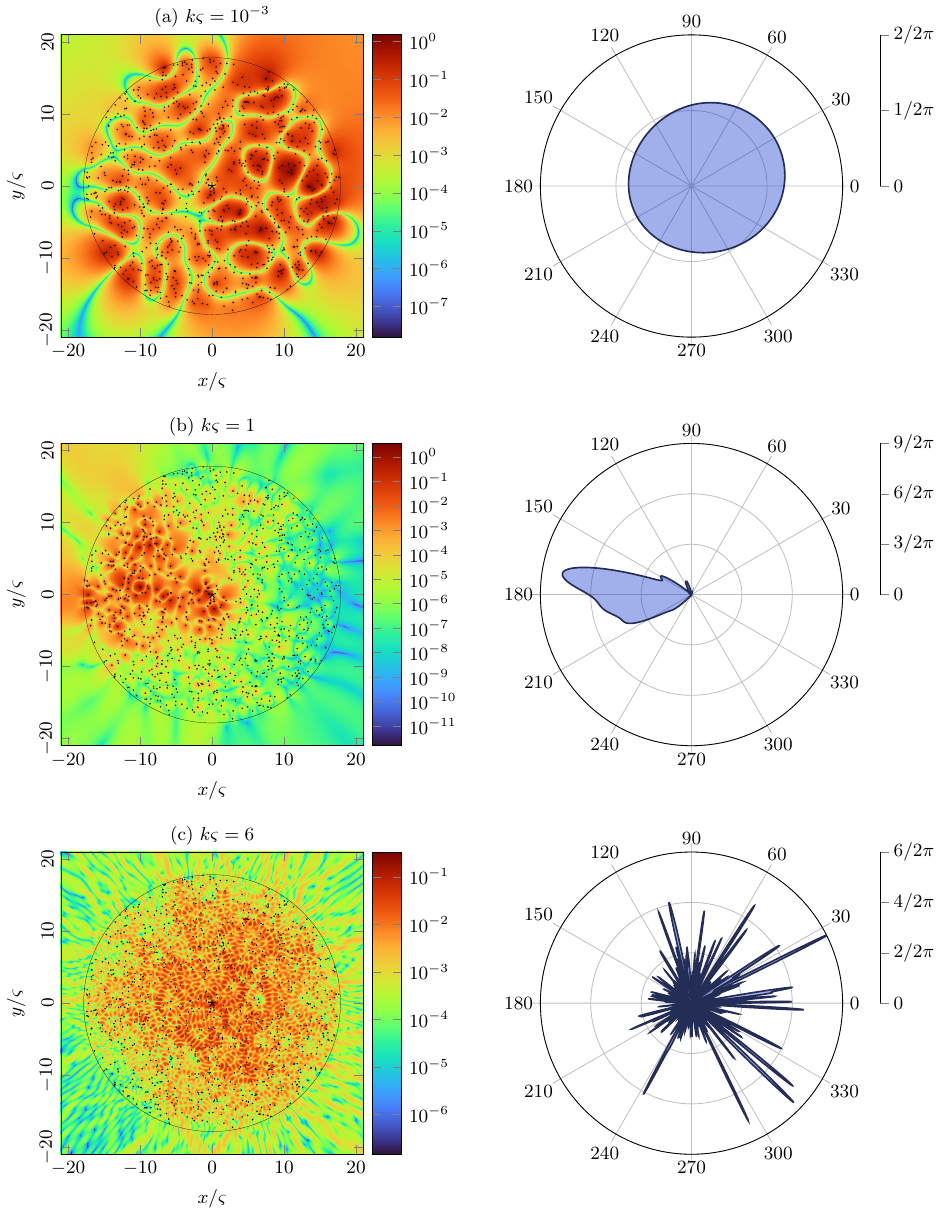}%
    \caption{\textbf{(left)} Square modulus of the full wavefunction $|\psi(k, \vvec{r} | \vvec{0})|^2$ and \textbf{(right)} angular PDF $\dd P(\theta) / \dd \theta$, for a wave propagating through a Lorentz gas of $N=1000$ scatterers. The gaseous configuration $\{\vvec{x}_1, \vvec{x}_2, ..., \vvec{x}_{1000}\}$ is the same along the three vertical panels. Panel (a): $k \varsigma = 10^{-3}$ and $\vvec{w}^2 \approx 0.012$, panel (b): $k \varsigma = 1$ and $\vvec{w}^2 \approx 0.83$, panel (c): $k \varsigma = 6$ and $\vvec{w}^2 \approx 0.04$. The directionality is very pronounced on panel (b). As $k \varsigma$ increases, the frequency content of $\dd P(\theta) / \dd \theta$ increases, as it goes from being almost isotropic and dominated by the $0$-th order Fourier term in panel (a), to being highly oscillatory in panel (c). This behavior is explained in \cref{appsec:fourier_analysis}.}%
    \label{fig:wave_functions}%
\end{figure*}%

\par
\begin{figure*}[t]%
    \hspace{-0.7cm}\includegraphics{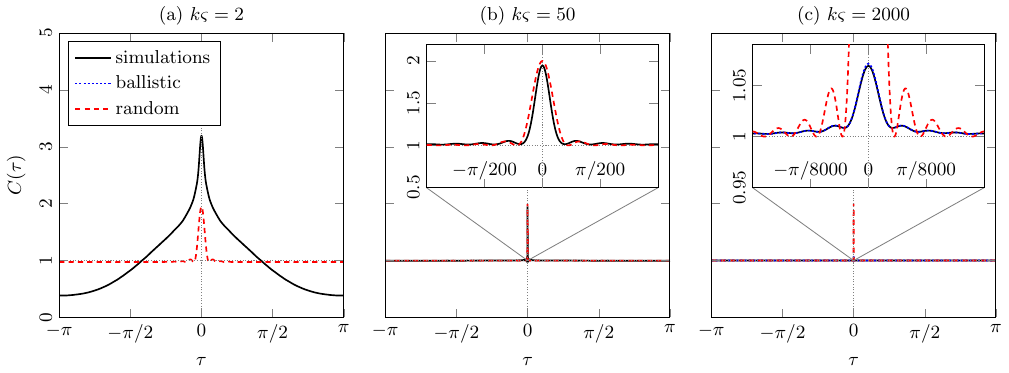}%
    \caption{Autocorrelation function $C(\tau)$ of the angular PDF $\dd P(\theta) / \dd \theta$, for a spherical wave propagating through a Lorentz gas of $N=1000$ scatterers. Panel (a): $k \varsigma = 2$, panel (b): $k \varsigma = 50$, panel (c): $k \varsigma = 2000$. In all three panels, the black curve represents the average over respectively $2000$, $800$ and $40$ random configurations of the gas, and the dashed red curve is the corresponding prediction for $C(\tau)$ associated with the random model of \cref{eq:random_model_correlation_function}. In panel (c), the dotted blue curve represents the analytical prediction in the ballistic regime from \cref{eq:correlation_function_delta_ballistic_2d}.}%
    \label{fig:autocorrelation_function}%
\end{figure*}

 To better understand the nature of this directionality peak, \cref{fig:wave_functions} presents the evolution of the full wavefunction $|\psi(k, \vvec{r} | \vvec{0})|^2$ and the angular PDF $\dd P(\theta) / \dd \theta$ for a wave propagating through a Lorentz gas of $N=1000$ scatterers with three different values of $k \varsigma$. The gaseous configuration is the same for all three panels, and was specifically chosen because of its high directionality for intermediate values of $k \varsigma$. For very low values of $k \varsigma$, when the wavelength is several orders of magnitude larger than the interscatterer distance, the gas acts as an effective medium, as the wave cannot spatially resolve its internal structure. The effective wavenumber is thus given by  \cite{Sheng2006_Introduction,Akkermans2007_Mesoscopic,Gaspard2024_Effective}
\begin{equation}
    k_{\text{eff}} = \rre \sqrt{k^2 - n F(k)},
\end{equation}
where $n$ is the density of the gas. For the situation shown in \cref{fig:wave_functions}(a), this leads to an effective wavelength $\lambda_{\text{eff}} \approx 6.5 \varsigma$ that is several orders of magnitude smaller than the incident wavelength $\lambda$, and which is in good agreement with the observed spatial period of the wavefunction. In the same panel, the angular PDF is almost isotropic, as it is largely dominated by the monopole term when $kR \ll 1$ (see \cref{appsec:fourier_analysis}). This explains the low value of the squared directionality $\vvec{w}^2$ observed in this regime. The second panel of the figure shows the wavefunction and the angular PDF in the same gaseous configuration, for $k \varsigma = 1$. In this situation, the wave is strongly localized towards the left direction. As a consequence, the angular PDF is highly directional ($\vvec{w}^2 \approx 0.83$), and the probability of detecting the wave in the right direction is almost zero. Even though this situation is among the most directional ones found during the simulations, \cref{fig:w_vs_ksigma} shows that, in this regime, many configurations lead to a non-negligible value of $\vvec{w}^2$. In this region, the autocorrelation function $C(\tau)$ is completely dominated by long-range correlations, as visible in \cref{fig:autocorrelation_function}(a). Finally, panel (c) of \cref{fig:wave_functions} shows the wavefunction and the angular PDF for $k \varsigma = 6$, just after the directionality peak. In this regime, there are numerous radial filaments of high intensity going out of the gas, in apparently random directions. This creates a highly oscillating angular PDF with no global preferred direction ($\vvec{w}^2 \approx 0.04$).

\subsection{Ballistic regime}\label{subsec:ballistic_regime}
\par In the ballistic regime, defined by the condition $\ell \gg 2R$ satisfied for large enough wavenumbers, the incident quantum particle undergoes a single scattering event before escaping the medium, and the first-order Born approximation is expected to hold. In this framework, the Foldy-Lax matrix defined in \cref{eq:foldy_lax_multiple_scattering_matrix} is approximated by \cite{Gaspard2022_Multiple}
\begin{equation}
    M_{ij} = F(k)^{-1} \delta_{ij},
\end{equation}
where the off-diagonal terms, resulting from multiple scattering events are neglected. The scattering amplitudes, which are solutions of \cref{eq:foldy_lax_linear_system}, are thus analytically given by
\begin{equation}\label{eq:born_approximation_amplitudes}
    a_j = F(k) G^+(k, \vvec{x}_j | \vvec{x}_0).
\end{equation}
Based on this, the autocorrelation function $C(\tau)$ is computed in \cref{appsec:autocorrelation_ballistic}. In two dimensions, its expression boils down to
\begin{equation}\label{eq:correlation_function_delta_ballistic_2d}
    C(\tau) = 1 + \frac{2}{k  \ell} \int_0^{kR} \! \dd z \, J_0(q z),
\end{equation}
where $q = 2 \sin(\tau/2)$, and the prefactor $\mu_0^2(k) / \mu^2(k, \vvec{x}_0)$ boils down to $1$ in virtue of \cref{eq:lloyd_ballistic,eq:single_scattering_amplitude_and_cross_section}. This function is shown in \cref{fig:autocorrelation_function}(c), and provides as expected a very good approximation of the actual autocorrelation function computed from simulations. The averaged directionality $\langle \vvec{w}^2 \rangle$ can then be computed from \cref{eq:correlation_function_delta_ballistic_2d,eq:directionality_squared_2}, yielding
\begin{align}
    \langle \vvec{w}^2 \rangle &= \frac{2}{\pi k \ell} \int_0^{\pi} \! \dd \tau \cos \tau \int_0^{kR} \! \dd z \, J_0(q z), \nonumber \\
    &= \frac{2}{k \ell} \int_0^{kR} \! \dd z \, J_1^2(z), \nonumber \\
    &= \frac{(kR)^3}{6 k \ell} \twoFthree{3/2}{3/2}{2}{5/2}{3}{-(kR)^2},
\end{align}
where the second line is found by inverting the integral order and using the fact that $\int_0^{\pi/2} \! \dd \theta \, \cos (2 n \theta) \, J_0(2 z \sin\theta)$ $=$ $\frac{\pi}{2} \, J_n^2(z)$ \cite{DLMF}, and the third line is a known integral form of the hypergeometric function $_2F_3$. As $kR \rightarrow +\infty$, this expression tends to \cite{Mathematica}
\begin{equation}\label{eq:ballistic_regime_w}
    \langle \vvec{w}^2 \rangle \xrightarrow[]{kR\to\infty} 2 \frac{\ln(kR) + \gamma + 3\ln(2) - 2}{\pi k \ell},
\end{equation}
where $\gamma \approx 0.577$ is the Euler-Mascheroni constant. This analytical prediction is compared to numerical simulations in \cref{fig:w_vs_ksigma}. As expected, the agreement is excellent in the ballistic regime $\ell \gg 2R$. 

\subsection{Random model of the directionality}\label{subsec:random_model}
\par To separate the contribution of the multiscattering from the intrinsic properties of a random angular PDF, the present section proposes a rudimentary model, based on which a simple expression for the directionality can be derived. As shown in \cref{appsec:fourier_analysis}, the angular PDF can be expressed as the square modulus of a Fourier expansion whose coefficients exponentially decay beyond a given frequency threshold dependent on $k$. On this basis, the proposed model assumes that
\begin{equation}\label{eq:random_model_pdf}
    \frac{\dd P}{\dd \theta}(\theta) = \frac{1}{2 \pi} \left| \sum_{s=0}^{S} V_s e^{\ii  s\theta} \right|^2,
\end{equation}
where the complex coefficients $V_s$ are the Fourier coefficients of the scattering amplitude $A(\theta)$. In the random model, the vector $\vvec{V} = (V_0, V_1, ..., V_S)$ is assumed to be randomly distributed over the unit sphere $\mathbb{C}^{S+1}$ in order to ensure proper normalization. As seen in \cref{appsec:fourier_analysis}, the approximate cutoff order at which the Fourier coefficients $V_s$ start to decay is given by $S \approx kR$.

\par \cref{eq:average_angular_pdf_random_model} shows that the random model satisfies the spherical symmetry criterion given by \cref{eq:isotropic_average_pdf}. It follows that the autocorrelation function can be computed from \cref{eq:angular_autocorrelation_function_spherical_case}, leading to
\begin{align}
    C(\tau) 
    &= \sum_{s,s',t,t'=0}^{S} \left\langle V_s V_{s'}^* V_t V_{t'}^* \right\rangle e^{\ii (s-s') \theta} e^{\ii (t-t') \theta'}, \nonumber \\
    &= \frac{(S+1)}{(S+2)} \left[ 1 + \frac{1}{(S+1)^2}\left|\sum_{s=0}^{S} e^{\ii s \tau}\right|^2 \right],
\end{align}
where the value of $\left\langle V_s V_{s'}^* V_t V_{t'}^* \right\rangle$ was computed from \cref{eq:fourth_order_moment_V}. The second term appearing in the brackets is proportional to the $(S+1)$-th Fejér kernel, and can be expressed in a closed form, yielding \cite{Stein2003_Fourier}
\begin{equation}\label{eq:random_model_correlation_function}
    C(\tau) = \frac{(S+1)}{(S+2)} \left\{ 1 + \frac{\sin^2\left[(S+1)\tau/2\right]}{(S+1)^2\sin^2(\tau/2)} \right\},
\end{equation}
which is shown in \cref{fig:autocorrelation_function} in red dashed curves. The random model provides a very good approximation of the actual autocorrelation function in the intermediate regime $1 < k\ell < 2kR$, as visible in \cref{fig:autocorrelation_function}(b). In the ballistic regime ($k\varsigma = 2000$), the oscillation of the autocorrelation function is also correctly predicted by the model, but their amplitude is highly overestimated. Finally, in the localized regime ($k\varsigma = 2$), the model misses the long-range correlations observed in the simulations.

\par The average squared directionality can finally be computed by injecting \cref{eq:random_model_correlation_function} into \cref{eq:directionality_squared_2}. This is done in \cref{appsubsec:directionality_random_model}, and yields
\begin{equation}\label{eq:random_model_w}
    \langle \vvec{w}^2 \rangle = \frac{S}{(S+1)(S+2)}.
\end{equation}
This result provides the average squared directionality for each value of $S \in \mathbb{N}$. As explained in \cref{appsec:fourier_analysis}, in the context of the Foldy-Lax model, the angular PDF is expected to contain Fourier terms up to $S \approx kR$. In \cref{fig:w_vs_ksigma}, the red dashed curves represent the squared directionality given by \cref{eq:random_model_w}, where $S$ has been treated as a continuous variable. We see that it predicts the global bell-shaped behavior of the directionality. In particular, it predicts its plateau $\langle \vvec{w}^2 \rangle \approx 0.2$, and the drift of the left flank of the curve towards smaller wavenumbers as $N$ increases. The latter phenomenon is due to the fact that the maximum of \cref{eq:random_model_w} occurs at $k=\sqrt{2}/R$ and that this value decreases as $N$ increases since the gas density is kept constant. This simple model, however, does not predict the strong directionality peak around $k \varsigma = 2$, nor does it provide the correct asymptotic behavior of the squared directionality for small and large wavenumbers. Indeed, it captures the statistical behavior of a generic angular distribution $\dd P(\theta) / \dd \theta$, without taking into account the underlying wave physics. In particular, this model ignores the long-range angular correlations \cite{Akkermans2007_Mesoscopic}, clearly visible in \cref{fig:autocorrelation_function}(a), which largely contribute to directionality according to \cref{eq:directionality_squared_2}.%

\section{Conclusion}\label{sec:conclusion}
\par In this article, we have established an observable, directionality $\vvec{w}$, to characterize the influence of the positions of individual scatterers in a Lorentz gas, on the presence probability density of a quantum particle emanating from its center. We have shown that, for a spherically symmetric gas, the directionality of the quantum particle as it exits the gas is much more significant than what is classically expected for a completely random angular distribution. The norm of the directionality vector even reaches a maximum at the limit of the Ioffe-Regel criterion ($k\ell = 1$), which shows that this observable is very sensitive to Anderson localization effects. As we have shown in the particular case of a single scatterer, this peak corresponds to a situation of maximum backscattering between nearby scatterers. This type of configuration is known to favor localization \cite{Akkermans2007_Mesoscopic,Sheng2006_Introduction}.

\par Further investigations are still needed to quantitatively explain the origin of this exceptionally high directionality in the multiple-scattering regime ($k \ell \gg 1$, $\ell \ll 2R$), and in the localized regime ($k\ell \lesssim 1$), and to characterize the asymptotic behavior of $\vvec{w}^2$ at small wavenumber ($kR \ll 1$). An important outcome of the present work is the identification of a direct connection between strong directionality and the presence of long-range angular correlations in the angular PDF, as shown in \cref{fig:autocorrelation_function}(a). This suggests that future theoretical descriptions of directionality should explicitly account for the mechanisms responsible for such correlations. 

\par In addition, since our present results are limited to two dimensions, an important next step is to generalize them to three dimensions, which is directly possible thanks to the flexibility of the Foldy-Lax formalism. Given the weaker impact of localization in three dimensions, we will probably need to consider larger Lorentz gases. To do this, more efficient numerical methods such as the “fast multipole method” (FMM) \cite{Greengard1987_Algorithm,Rokhlin1990_Rapid,Lu1993_Algorithm,Cho2010_Wideband} could be used to substantially increase the number of scatterers ($N \approx 10^6$) and determine whether directionality remains as important for very large systems. Moreover, more complex interactions accounting for the finite size of the scatterers could be tested, following the Kohn-Korringa-Rostoker \cite{Kohn1954_Solution,Korringa1947_Calculation,Korringa1994_Early} or Faddeev-Watson \cite{Faddeev1961_Scattering,Joachain1987_Quantum,Joachain2023_Multiple} approaches.

\par Coming back to our original motivation to study the quantum measurement problem, we aim at progressively making our model more realistic, with a view to experimental situations involving collisional decoherence. Following Mott's original idea, it would be interesting to investigate the emergence of directionality in quantum models that explicitly include the detector degrees of freedom in the full wavefunction, for which collisional quantum decoherence may play an important role. This could be done by first allowing for scatterer (atom/molecule) excitations, then by explicitly taking the electron recoil into account, and finally by describing the global entanglement of the quantum particle with the detector electrons or atoms. Depending on the numerical complexity of these models, they could be combined again with statistical approaches \cite{Gaspard2023_Transverse,Gaspard2022_master}, thus providing a framework to investigate to what extent, at least within such models, the deterministic Schrödinger equation alone can lead to both directionality and quantum decoherence.%

\begin{acknowledgments}
    \par This work was supported by the Belgian Fonds de la Recherche Scientifique -- FNRS under IISN Grant No. 4.45.10.08. This research has also been supported by the ANR project MARS\_light under reference \href{https://anr.fr/Projet-ANR-19-CE30-0026}{ANR-19-CE30-0026}, by the program “Investissements d'Avenir" launched by the French Government. It also received support from a grant of the Simons Foundation (No. 1027116).
\end{acknowledgments}

\appendix%

\section{Squared directionality and autocorrelation function}\label{appsec:autocorrelation}
\par In this appendix, we prove the equivalence between \cref{eq:directionality_squared,eq:directionality_squared_2} of the main text, and we further discuss the normalization of the autocorrelation function $C(\tau)$.

\par The double integration appearing in \cref{eq:directionality_squared} can be simplified by first defining the angle $\tau$ as the angle between $\vvec{\Omega}$ and $\vvec{\Omega}'$, such that
\begin{equation}
    \cos(\tau) \defEqual \vvec{\Omega} \cdot \vvec{\Omega}'.
\end{equation}
This allows to express the direction $\vvec{\Omega'}$ as
\begin{equation}
    \vvec{\Omega}' = \cos(\tau) \vvec{\Omega} + \sin(\tau) \vvec{u},
\end{equation}
where $\vvec{u}$ is a unit vector orthogonal to $\vvec{\Omega}$. The integral over $\vvec{\Omega}'$ can thus be separated into an integral over $\tau$ and an integral over $\vvec{u}$ as
\begin{equation}\label{eq:decomposition_omega_prime}
    \oint_{\mathcal{S}_d} \dd \Omega' = \int_{0}^{\pi} \dd \tau \, \sin^{d-2} (\tau) \oint_{\mathcal{S}_{d-1}} \dd u.
\end{equation}
Using this decomposition and reorganizing the integrals, the expression of $\langle \vvec{w}^2 \rangle$ in \cref{eq:directionality_squared} becomes
\begin{equation}\label{eq:directionality_squared_2_appendix}
    \langle \vvec{w}^2 \rangle = \frac{S_{d-1}}{S_d} \int_{0}^{\pi} \dd \tau \, \sin^{d-2} (\tau) \cos(\tau) C(\tau) 
\end{equation}
where the correlation function $C(\tau)$ is defined as
\begin{align}\label{eq:correlation_function_definition}
    C(\tau) \ & \defEqual \frac{S_d}{S_{d-1}} \oint_{\mathcal{S}_d} \dd \Omega \oint_{\mathcal{S}_{d-1}} \dd u \nonumber \\
    & \times \left\langle \frac{\dd P}{\dd \Omega}(\vvec{\Omega}) \frac{\dd P}{\dd \Omega}(\cos(\tau) \vvec{\Omega} + \sin(\tau) \vvec{u}) \right\rangle.
\end{align}
\cref{eq:directionality_squared_2_appendix} corresponds to \cref{eq:directionality_squared_2} of the main text. The normalization factors chosen above are such that the background correlation value $C_\text{back}$, given for the special case of a constant angular PDF, is equal to $1$. In the special case of a spherically symmetric problem, for which the disordered average angular distribution has the constant value
\begin{equation}\label{eq:isotropic_average_pdf}
    \left\langle \frac{\dd P}{\dd \Omega}(\vvec{\Omega}) \right\rangle = \frac{1}{S_d},
\end{equation}
the correlator appearing in \cref{eq:correlation_function_definition} is independent of $\vvec{\Omega}$ and $\vvec{u}$. By extracting it from the integrals, one gets
\begin{equation}\label{eq:correlation_function_spherical_case}
    C(\tau) = S_d^2 \left\langle \frac{\dd P}{\dd \Omega}(\vvec{\Omega}) \frac{\dd P}{\dd \Omega}(\vvec{\Omega}') \right\rangle,
\end{equation}
where $\vvec{\Omega}$ and $\vvec{\Omega}'$ are two arbitrary directions separated by the angle $\tau$.

\par By squaring \cref{eq:normalization_condition} and averaging it over the disorder, one gets the relation
\begin{equation}
    \oint_{\mathcal{S}_d} \dd \Omega \oint_{\mathcal{S}_d} \dd \Omega' \, \left\langle \frac{\dd P}{\dd \Omega}(\vvec{\Omega}) \frac{\dd P}{\dd \Omega}(\vvec{\Omega}') \right\rangle = 1,
\end{equation}
which can be further simplified by separating the integral over $\vvec{\Omega}'$ into an integral over $\tau$ and an integral over $\vvec{u}$ via \cref{eq:decomposition_omega_prime}, as done previously. This leads to the normalization condition for the correlation function, which reads
\begin{equation}\label{eq:correlation_function_normalization}
    \frac{S_{d-1}}{S_d} \int_{0}^{\pi} \dd \tau \, \sin^{d-2} (\tau) C(\tau) = 1.
\end{equation}%

\section{Derivation of the integrated probability currents}\label{appsec:probability_currents}
\par This appendix is devoted to the computation of the flux integrals appearing in \cref{eq:conservation_integral}, which allow us to retrieve \cref{eq:normalization_C}.

\par In the far-field region $\partial \mathcal{B}_\infty$, combining the definition of the probability current and \cref{eq:scattering_amplitude_definition}, one gets that $\vvec{J}(\vvec{r})$ asymptotically behaves as
\begin{align}
    \vvec{J}(\vvec{r}) \ & \xrightarrow[]{r \rightarrow \infty} k \vvec{\Omega} \, |G^+(k, \vvec{r} | \vvec{x}_0)|^2 \, |A(\vvec{\Omega})|^2, \nonumber \\
    & \xrightarrow[]{r \rightarrow \infty} \vvec{\Omega} \, \frac{\pi \mu_0(k)}{S_d r^{d-1}} \, |A(\vvec{\Omega})|^2,
\end{align}
where the transition to the second line is due to the asymptotic behavior of the free Green function recalled in \cref{eq:free_green_asymptotic_2}. The integral over the surface $\partial \mathcal{B}_\infty$ can thus be written
\begin{equation}\label{eq:I_infinity}
    I_\infty = \frac{\pi \mu_0(k)}{S_d} \oint_{\mathcal{S}_d} \dd \Omega \, |A(\vvec{\Omega})|^2.
\end{equation}

\par On the other hand, in the direct vicinity of the emitter $\partial \mathcal{B}_0$, the divergence theorem applied to the integral $I_0$ gives
\begin{align}
    I_0 \ & = \int_{\mathcal{B}_0} \dd \vvec{r} \, \vvec{\nabla}_{\vvec{r}} \cdot \vvec{J}(\vvec{r}), \nonumber \\
    & = \int_{\mathcal{B}_0} \dd \vvec{r} \, \vvec{\nabla}_{\vvec{r}} \cdot \iim [\psi^-(k, \vvec{r} | \vvec{x}_0) \vvec{\nabla}_{\vvec{r}} \psi^+(k, \vvec{r} | \vvec{x}_0)], \hspace{.5cm}
\end{align}
which, using the relation $\vvec{\nabla}_{\vvec{r}} \cdot (f \vvec{\nabla}_{\vvec{r}} g - g \vvec{\nabla}_{\vvec{r}} f) = f \nabla^2_{\vvec{r}} g - g \nabla^2_{\vvec{r}} f$, becomes
\begin{align}\label{eq:I_zero}
    I_0 \ & = \int_{\mathcal{B}_0} \dd \vvec{r} \, \iim [\psi^-(k, \vvec{r} | \vvec{x}_0) \nabla^2_{\vvec{r}} \psi^+(k, \vvec{r} | \vvec{x}_0)], \nonumber \\
    & = \int_{\mathcal{B}_0} \dd \vvec{r} \, \iim  \left[ \psi^-(k, \vvec{r} | \vvec{x}_0) \delta^{(d)}(\vvec{r} - \vvec{x}_0) \right], \nonumber \\
    & = - \iim  \left[ \psi^+(k, \vvec{x}_0 | \vvec{x}_0) \right], \nonumber \\
    & = \pi \mu(k, \vvec{x}_0),
\end{align}
where the second line is due to the Schrödinger equation, and the last line comes from \cref{eq:full_local_density_of_states}.%

\section{Computation of the directionality vector in the Foldy-Lax model}\label{appsec:analytical_w}
\par In this appendix, we derive the analytical expression of the directionality vector in the Foldy-Lax framework, given in \cref{eq:analytical_foldy_lax_w_final_form}. It is done by firstly injecting \cref{eq:foldy_lax_pdf} in \cref{eq:directionality_def}. This gives
\begin{align}\label{eq:analytical_w_foldy_lax_int}
    \vvec{w} \ & = \frac{\mu_0(k)}{\mu(k, \vvec{x}_0)} \oint_{\mathcal{S}_d} \frac{\dd \Omega}{S_d} \left| \sum_{j=0}^N a_j e^{-\ii  k \vvec{\Omega} \cdot (\vvec{x}_j - \vvec{x}_0)} \right|^2 \vvec{\Omega}, \nonumber \\
    & = \frac{2 \mu_0(k)}{\mu(k, \vvec{x}_0)} \rre \left[ \sum_{l, j<l}^N a_j a_l^* \oint_{\mathcal{S}_d} \frac{\dd \Omega}{S_d} \vvec{\Omega} e^{\ii  k \vvec{\Omega} \cdot (\vvec{x}_l - \vvec{x}_j)}\right], \hspace{.5cm}
\end{align}
where the constant term resulting from the square modulus of the sum vanishes as it is integrated over the sphere, and where we defined $a_0=1$ for the sake of readability. The integral appearing in the right-hand side of this equation can be exactly expressed by using the known relation \cite{Gaspard2022_Multiple,Grafakos2010_Classical}
\begin{equation}\label{eq:integral_plane_wave}
    \oint_{\mathcal{S}_d} \frac{\dd \Omega}{S_d} e^{\ii  \vvec{\Omega} \cdot \vvec{z}} = \Gamma\left( \frac{d}{2} \right) \left( \frac{z}{2} \right)^{1-\frac{d}{2}} J_{\frac{d}{2}-1}(z),
\end{equation}
where $\vvec{z} = z \vvec{e}_z$. Differentiating this expression with respect to $z$ leads to \cite{DLMF}
\begin{equation}
    \vvec{e}_z \cdot \oint_{\mathcal{S}_d} \frac{\dd \Omega}{S_d} \vvec{\Omega} e^{\ii  \vvec{\Omega} \cdot \vvec{z}} = \ii \Gamma\left( \frac{d}{2} \right) \left( \frac{z}{2} \right)^{1-\frac{d}{2}} J_{\frac{d}{2}}(z).
\end{equation}
By symmetry, the vector integral appearing in the left-hand side of this equation is collinear to $\vvec{e}_z$, and can thus be expressed as
\begin{align}\label{eq:integral_Z}
    \oint_{\mathcal{S}_d} \frac{\dd \Omega}{S_d} \vvec{\Omega} e^{\ii  \vvec{\Omega} \cdot \vvec{z}} &= \ii \Gamma\left( \frac{d}{2} \right) \left( \frac{z}{2} \right)^{1-\frac{d}{2}} J_{\frac{d}{2}}(z) \vvec{e}_z, \nonumber \\
    &= \frac{\ii \vvec{z}}{d} \Gamma\left( \frac{d+2}{2} \right) \left( \frac{z}{2} \right)^{-\frac{d}{2}} J_{\frac{d}{2}}(z), \nonumber \\
    &= \frac{\ii \vvec{z}}{d} Z_d(z),
\end{align}
where the function $Z_d(z)$, defined for the occasion, is real-valued, and normalized by $Z_d(0) = 1$. Injecting this expression for $\vvec{z} = k (\vvec{x}_l - \vvec{x}_j)$ in \cref{eq:analytical_w_foldy_lax_int} leads to \cref{eq:analytical_foldy_lax_w_final_form} of the main text, that reads
\begin{align}
    \vvec{w} \ & = \frac{2 \mu_0(k)}{\mu(k, \vvec{x}_0)} \rre \left[ \sum_{l, j<l}^N a_j a_l^* \frac{\ii  k (\vvec{x}_l - \vvec{x}_j)}{d} Z_d(k \Norm{\vvec{x}_l - \vvec{x}_j}) \right], \nonumber \\
    & = \frac{2k}{d} \frac{\mu_0(k)}{\mu(k, \vvec{x}_0)} \sum_{l, j<l}^N \iim (a_l a_j^*) (\vvec{x}_l - \vvec{x}_j) Z_d(k \Norm{\vvec{x}_l - \vvec{x}_j}), \nonumber  \\
\end{align}%

\section{Fourier analysis of the angular PDF in two dimensions}\label{appsec:fourier_analysis}
\par This appendix is devoted to the analysis of the Fourier expansion of the spherical scattering amplitude $A(\vvec{\Omega})$ and of the angular PDF $\dd P(\theta) / \dd \theta$ in two dimensions. The former Fourier representation is found by inserting the Jacobi-Anger expansion \cite{DLMF} in \cref{eq:foldy_lax_scattering_amplitude}, yielding
\begin{align}
    A(\theta) &= 1 + \sum_{s=-\infty}^{+\infty} e^{\ii s \theta} \left[ \ii^s \sum_{j=1}^N a_j J_s(k r_j) e^{-\ii s \theta_j} \right],
\end{align}
where $r_{j} = ||\vvec{x}_l - \vvec{x}_0||$ is the distance between the scatterer $j$ and the emitter, while $\theta_{j}$ is the angle between the vector $\vvec{x}_l - \vvec{x}_0$ and the $x$-axis. The presence of the Bessel function $J_s(k r_{j})$ in this expression implies that the contribution of a given scatterer $j$ to the Fourier coefficient is strongly damped when $s \gg k r_{j}$. This is due to the asymptotic expression of the Bessel function of the first kind for small arguments, which behaves as $(kr_{j})^s / (2^s s!)$ \cite{DLMF}. Therefore, a scatterer $j$ contributes significantly only to the Fourier coefficients of order $-kr_{j} < s < kr_{j}$. The complete Fourier series of $A(\theta)$ is thus limited to a finite number of terms, which can be deduced from the maximum value of $kr_{j}$ in the system. As the scatterers lie in a disk of radius $R$ centered on the emitter position, the maximum value for $r_j$ is $R$. Therefore, the Fourier series of $A(\theta)$ is approximately limited to the range 
\begin{equation}\label{eq:fourier_cutoff}
    |s| \leq S_\text{cutoff} = kR.
\end{equation}
As the angular PDF is proportional to $|A(\theta)|^2$, its Fourier series is approximately limited to the range $[-2kR, 2kR]$. This behavior is indeed qualitatively observed in \cref{fig:wave_functions}, where the angular PDF is seen to oscillate with an increasing frequency as $k\varsigma$ (and thus $kR$) increases. Following the Nyquist-Shannon sampling theorem \cite{Grafakos2010_Classical}, the angular fluctuations visible on that figure are given by
\begin{equation}\label{eq:angular_step_cutoff}
    \Delta \theta < \frac{2\pi}{S_\text{cutoff}} = \frac{\pi}{kR},
\end{equation}
which is in agreement with \refRef{Gaspard2022_Multiple}.%

\section{Autocorrelation function in the ballistic regime}\label{appsec:autocorrelation_ballistic}
\par The aim of this appendix is to compute the autocorrelation function $C(\tau)$ in the high-wavenumber ballistic regime. In this regime, the Born approximation consists in retaining only single-scattering contributions, which leads to a truncation of the expansion in the scattering amplitude at lowest order. We consider here a disk-shaped gas, where the scatterers are uniformly and independently distributed inside a ball $B_R(\vvec{x}_0)$ of radius $R$ centered at the emitter position $\vvec{x}_0$. In this case, the single-scatterer spatial probability density is $\rho = 1/V_R$, where $V_R$ is the volume of the medium. The symmetry of this system ensures that \cref{eq:isotropic_average_pdf} is satisfied, as explained in \cref{subsec:multiple_particles}. Therefore, the autocorrelation function $C(\tau)$ can be computed in the ballistic regime by injecting \cref{eq:foldy_lax_pdf} into \cref{eq:correlation_function_spherical_case}. Upon keeping terms up to the second order in amplitude, one gets
\begin{align}
    C(\tau) &= \frac{\mu_0^2(k)}{\mu^2(k,\vvec{x}_0)} \left\{1 + 2 \rre \sum_{j=1}^N [\langle A_j(\vvec{\Omega}) \rangle + \langle A_j(\vvec{\Omega'}) \rangle]\right. \nonumber \\
    & + \sum_{j,l=1}^{N} [\langle A_j(\vvec{\Omega}) A_l^*(\vvec{\Omega}) \rangle + \langle A_j(\vvec{\Omega'}) A_l^*(\vvec{\Omega'}) \rangle] \nonumber \\
    & \left.+ 2 \rre\sum_{\substack{j,l=1}}^N [\langle A_j(\vvec{\Omega}) A_l(\vvec{\Omega'}) \rangle + \langle A_j(\vvec{\Omega}) A_l^*(\vvec{\Omega'}) \rangle] \right\}.
\end{align}
In this expression, the ratio $\mu_0^2(k)/\mu^2(k,\vvec{x}_0)$ has been factored out of the disorder-averages because it does not depend on the gas configuration, as can be for example verified from the Lloyd formula, which reads in the ballistic regime \cite{Gonis2000_Multiple}
\begin{equation}\label{eq:lloyd_ballistic}
    \mu(k, \vvec{x}_0) - \mu_0(k) = \frac{N}{2 \pi k} \iim \left[ \frac{\dd}{\dd k} \ln F(k) \right].
\end{equation}

\par Under the assumption that the obstacles are identically and independently distributed, the mixed moments can be factorized when $j \neq l$, and any moment depending only on one scatterer index $j$ becomes independent of that index. In addition to that, the sphericity of the system ensures that $\langle A_j(\vvec{\Omega}) \rangle$, $\langle A_j(\vvec{\Omega})^2 \rangle$, and $\langle |A_j(\vvec{\Omega})|^2 \rangle$ are independent of the direction. This allows to further simplify the expression of the correlation function as
\begin{align}\label{eq:correlation_function_ballistic_general}
    C(\tau) &= \frac{\mu_0^2(k)}{\mu^2(k,\vvec{x}_0)} \left\{1 + 4 N \rre \langle A_1(\vvec{\Omega}) \rangle + 2 N \langle |A_1(\vvec{\Omega})|^2 \rangle \right. \nonumber \\
    &+ 4 N(N-1) |\langle A_1(\vvec{\Omega}) \rangle|^2 + 2 N(N-1) \rre [\langle A_1(\vvec{\Omega}) \rangle^2]  \nonumber \\
    & \left. + 2 N \rre \langle A_1(\vvec{\Omega}) A_1(\vvec{\Omega'}) \rangle + 2 N \rre \langle A_1(\vvec{\Omega}) A_1^*(\vvec{\Omega'}) \rangle \right\},
\end{align}
where only the last two terms depend on the angle $\tau$. 

\par The moment $\langle A_1(\vvec{\Omega}) \rangle$ can be computed via the volume integral, which generalizes \cref{eq:foldy_lax_scattering_amplitude} to a continuum of scatterers,
\begin{equation}
    \langle A_1(\vvec{\Omega}) \rangle = \int_{B_R(\vvec{x}_0)} \dd \vvec{x} \, \rho \, a(\vvec{x}) e^{-\ii k \vvec{\Omega} \cdot (\vvec{x} - \vvec{x}_0)},
\end{equation}
where $a(\vvec{x})$ is the scattering amplitude of a scatterer located at position $\vvec{x}$. Injecting the expression of $a(\vvec{x})$ from \cref{eq:born_approximation_amplitudes} leads to
\begin{align}
    \langle A_1(\vvec{\Omega}) \rangle &= \frac{\rho F(k)}{4 \ii} \left(\frac{k}{2 \pi}\right)^{\nu} \int_{0}^{R} \dd r \, r^{\nu+1} H^{+}_{\nu}(k r) \nonumber \\
    & \times \oint_{\mathcal{S}_d} \dd \Omega \, e^{-\ii k \vvec{\Omega} \cdot (\vvec{x} - \vvec{x}_0)},
\end{align}
where $\nu = d/2 - 1$ and $r = \Norm{\vvec{x} - \vvec{x}_0}$. The angular integral present in the above expression can be explicitly computed from \cref{eq:integral_plane_wave}. Therefore, the sought moment becomes
\begin{equation}\label{eq:average_Aj_intermediate}
    \left\langle A_1(\vvec{\Omega}) \right\rangle = \frac{\pi \rho F(k)}{2 \ii} \int_{0}^R \dd r \, r H^{+}_{\nu}(k r) J_{\nu}(k r).
\end{equation}
The radial integral appearing in this expression has the closed form expression \cite{DLMF}
\begin{align}
    I = \frac{R^2}{2} \left[ H^+_{\nu}(kR) J_{\nu}(kR) - H^+_{\nu-1}(kR) J_{\nu+1}(kR) \right],
\end{align}
which, following the asymptotic behavior of the Bessel and Hankel functions for large arguments \cite{DLMF}, tends to $R/(\pi k)$ when $kR \gg 1$. Therefore, in the ballistic regime and for a large medium, the average value of the amplitude becomes
\begin{equation}\label{eq:average_Aj_final}
    \left\langle A_1(\vvec{\Omega}) \right\rangle = \frac{\rho R F(k)}{2 \ii k},
\end{equation}
which is independent on the direction $\vvec{\Omega}$ and on the scatterer index, as expected.

\par Following similar steps, the moment $\langle A_1(\vvec{\Omega}) A_1^*(\vvec{\Omega'}) \rangle$ can be expressed as
\begin{align}\label{eq:moment_A1_A1_conjugate_intermediate}
    \langle A_1(\vvec{\Omega}) A_1^*(\vvec{\Omega'}) \rangle &= (2\pi)^{\nu+1} \frac{\rho |F(k)|^2 }{(qk)^{\nu}} \nonumber \\
    & \times \int_{0}^{R} \dd r \, r^{\nu + 1} |G^+(k, \vvec{r} | \vvec{x}_0)|^2 J_{\nu}(q k r),
\end{align}
where $q = \Norm{\vvec{\Omega} - \vvec{\Omega'}} = 2 \sin (\tau/2)$. Injecting \cref{eq:free_green_asymptotic_2} in it, this expression becomes \cite{Prudnikov1986_Integrals}
\begin{align}\label{eq:moment_A1_A1_conjugate}
     \langle A_1(\vvec{\Omega}) A_1^*(\vvec{\Omega'}) \rangle &= \frac{(2\pi)^{\nu+1}}{S_d} \frac{\sigma \rho}{k} \int_{0}^{kR} \frac{\dd z}{(qz)^{\nu}} \, J_{\nu}(q z), \nonumber \\
     &= \sigma \rho R \, \oneFtwo{\frac{1}{2}}{\frac{3}{2}}{\nu+1}{-\frac{(qkR)^2}{4}},
\end{align}
where we used the fact that $|F(k)|^2 = k \sigma / \pi \mu_0(k)$ \cite{Gaspard2022_Multiple}. In particular, for $\tau = 0$, this moment reduces to
\begin{equation}\label{eq:moment_A1_A1_conjugate_zero}
    \langle |A_1(\vvec{\Omega})|^2 \rangle = \sigma \rho R.
\end{equation}
It is important to note that the transition from \cref{eq:moment_A1_A1_conjugate_intermediate} to \cref{eq:moment_A1_A1_conjugate} was performed by injecting the asymptotic behavior of the free Green function in the expression. Even though the integral also contains contributions from the near-field region $kr \lesssim 1$, the integrand remains finite in this region if $d \leq 3$, ensuring the asymptotic validity of this approximation.

\par The last moment that is present in \cref{eq:correlation_function_ballistic_general} is $\langle A_1(\vvec{\Omega}) A_1(\vvec{\Omega'}) \rangle$. Its expression is given by an integral similar to \cref{eq:moment_A1_A1_conjugate_intermediate}, but that contains a factor $G^+(kr)^2$. Its integrand is therefore highly oscillatory, and this integral decays much faster than the other as $kR \rightarrow +\infty$, which makes it negligible here.

\par The final expression of the autocorrelation function in the ballistic regime for a disk-shaped medium of radius $R$ can be gathered by injecting \cref{eq:average_Aj_final,eq:moment_A1_A1_conjugate,eq:moment_A1_A1_conjugate_zero} in \cref{eq:correlation_function_ballistic_general}. For $d=2$ and $3$, all constant terms become negligible in front of $1$ as $kR \rightarrow +\infty$. Therefore, the autocorrelation function reads
\begin{align}
    C(\tau) &= \frac{\mu_0^2(k)}{\mu^2(k,\vvec{x}_0)} \left[1 + 2 N \rre \langle A_1(\vvec{\Omega}) A_1^*(\vvec{\Omega'}) \rangle \right],\nonumber \\
    &= \frac{\mu_0^2(k)}{\mu^2(k,\vvec{x}_0)} \left[1 + \frac{2R}{\ell} \oneFtwo{\frac{1}{2}}{\frac{3}{2}}{\nu+1}{-\frac{(qkR)^2}{4}}\right],
\end{align}
where we used the relation $N \sigma \rho = 1/\ell$. This expression corresponds to \cref{eq:correlation_function_delta_ballistic_2d} of the main text, where it is particularized to $d=2$, and where the integral form of the hypergeometric function (which is shown in \cref{eq:moment_A1_A1_conjugate}) is used instead, to ease the calculations.%

\section{Derivations for the random model}\label{appsec:random_model}
\par This appendix provides detailed derivations that are needed for the random model presented in \cref{subsec:random_model}. 

\subsection{Complex moments of random vectors on the unit sphere}\label{appsubsec:complex_moments_unit_sphere}
\par In this appendix, we recall some properties of random vectors uniformly distributed on the unit sphere in $\mathbb{C}^n$. Throughout this appendix, $\vvec{V} = (V_1, V_2, \ldots, V_n) \in \mathbb{C}^n$ refers to such a random vector, thus satisfying the normalization condition
\begin{equation}
    \sum_{i=1}^{n} |V_i|^2 = 1,
\end{equation}
while $\vvec{Z} = (Z_1, Z_2, \ldots, Z_n) \in \mathbb{C}^n$ is a random vector following the standard complex normal distribution $\mathcal{CN}_n(\vvec{0},\vvec{I}_n)$, meaning that its components are independent and identically distributed along the complex normal distribution $\mathcal{CN}(0,1)$. Moreover, the real and imaginary parts of each of its components are independent and identically distributed along the real normal distribution $\mathcal{N}(0,1/2)$.

\par Because $\mathcal{CN}_n(\vvec{0},\vvec{I}_n)$ is isotropic, the random direction $\Norm{\vvec{Z}}^{-1} \vvec{Z}$ is independent on the norm $\Norm{\vvec{Z}}$, and is uniformly distributed along the unit sphere in $\mathbb{C}^n$ \cite{Mardia2000_Directional}. This property directly leads to
\begin{equation}\label{eq:V_Z_equality_distribution}
    \vvec{V} \overset{d}{=} \Norm{\vvec{Z}}^{-1} \vvec{Z} \quad\Leftrightarrow\quad \vvec{Z} \overset{d}{=} \Norm{\vvec{Z}} \vvec{V},
\end{equation}
where $\overset{d}{=}$ denotes equality in distribution. This equality could actually be achieved with any isotropic distribution, but $\mathcal{CN}_n(\vvec{0},\vvec{I}_n)$ is particularly convenient for calculations.

\par We try hereafter to compute the general moment $\left\langle \prod_{i=1}^{n} V_i^{p_i} V_i^{*q_i} \right\rangle$, where $p_i, q_i \in \mathbb{N}$ for all $i \in \{1, 2, \ldots, n\}$. To do so, we start by writing the equivalent moment for the vector $\vvec{Z}$ as \cite{Fassino2019_Computing}
\begin{equation}\label{eq:Z_moment_1}
    \left\langle \prod_{i=1}^{n} Z_{i}^{p_i} Z_{i}^{*q_i} \right\rangle = \prod_{i=1}^{n} \left\langle Z_{i}^{p_i} Z_{i}^{*q_i} \right\rangle = \prod_{i=1}^{n} \delta_{p_i,q_i} p_i!,
\end{equation}
which is non-zero only if $p_i = q_i$ for all $i$. The same moment can also be expressed in terms of the vector $\vvec{V}$ via \cref{eq:V_Z_equality_distribution}, yielding
\begin{align}\label{eq:Z_moment_2}
    \left\langle \prod_{i=1}^{n} Z_{i}^{p_i} Z_{i}^{*q_i} \right\rangle 
    &= \left\langle \Norm{\vvec{Z}}^{\sum_{i=1}^{n} (p_i + q_i)} \prod_{i=1}^{n} V_{i}^{p_i} V_{i}^{*q_i} \right\rangle, \nonumber \\
    &= \left\langle \Norm{\vvec{Z}}^{\sum_{i=1}^{n} (p_i + q_i)} \right\rangle \left\langle \prod_{i=1}^{n} V_{i}^{p_i} V_{i}^{*q_i} \right\rangle,
\end{align}
where we used the independence between the norm $\Norm{\vvec{Z}}$ and the direction $\vvec{V}$ to separate the averages. Before pursuing with the calculation of the first factor, it is useful to remember that, if $A_1, \ldots, A_m$ are $m$ independent random variables distributed along $\mathcal{N}(0,1)$, then $\sum_{i=1}^{m} A_i^2 \sim \text{Gamma}(m/2, 2)$  \cite{Krishnamoorthy2016_Handbook}. Therefore, as the squared norm $\Norm{\vvec{Z}}^2 = \sum_{i=1}^{n} |Z_i|^2$ is the sum of $2n$ independent squared variables distributed along $\mathcal{N}(0,1/2)$ (the real and imaginary parts of each $Z_i$, denoted $Y_j$ hereafter), its distribution follows
\begin{equation}
    \sum_{j=1}^{2n} Y_j^2 \overset{d}{=} \frac{1}{2} \sum_{j=1}^{2n} A_j^2 \sim \frac{1}{2} \text{Gamma}(n, 2) = \text{Gamma}(n, 1).
\end{equation}
The first unknown factor appearing in \cref{eq:Z_moment_2} can thus be computed as
\begin{align}\label{eq:Z_norm_moment}
     \left\langle \Norm{\vvec{Z}}^{\sum_{i=1}^{n} (p_i + q_i)} \right\rangle 
     &= \left\langle \left(\Norm{\vvec{Z}}^2\right)^{\sum_{i=1}^{n} (p_i + q_i)/2} \right\rangle, \nonumber \\
     &= \frac{\Gamma \left[ n+\sum_{i=1}^{n} (p_i + q_i)/2\right]}{\Gamma(n)},
\end{align}
where the second equality is a known property of the Gamma distribution \cite{Krishnamoorthy2016_Handbook}. Combining \cref{eq:Z_moment_1,eq:Z_moment_2,eq:Z_norm_moment}, the sought moment for the vector $\vvec{V}$ reads
\begin{equation}\label{eq:V_moment_final}
    \left\langle \prod_{i=1}^{n} V_i^{p_i} V_i^{*q_i} \right\rangle = \frac{\Gamma(n)}{\Gamma \left[ n+\sum_{i=1}^{n} (p_i + q_i)/2\right]} \prod_{i=1}^{n} \delta_{p_i,q_i} p_i!.
\end{equation}

\par In \cref{subsec:random_model}, the random variable $\vvec{V}$ has $n = S+1$ components. Therefore, the second-order moment of $\vvec{V}$ reads
\begin{equation}\label{eq:second_order_moment_V}
    \left\langle V_s V_t^* \right\rangle = \frac{\Gamma(S+1)}{\Gamma(S+2)} \delta_{s,t},
\end{equation}
while its fourth-order moment reads
\begin{equation}\label{eq:fourth_order_moment_V}
    \left\langle V_s V_{s'}^* V_t V_{t'}^* \right\rangle = \frac{\Gamma(S+1)}{\Gamma(S+3)} (\delta_{s,s'}\delta_{t,t'} + \delta_{s,t'}\delta_{t,s'}).
\end{equation}

\subsection{Average value of the angular PDF in the random model}\label{appsubsec:average_angular_pdf_random_model}
\par In the random model, the averaged angular distribution has the form
\begin{align}\label{eq:average_angular_pdf_random_model}
    \left\langle \frac{\dd P}{\dd \theta}(\theta) \right\rangle &= \frac{1}{2 \pi} \sum_{s,t=0}^{S} \left\langle V_s V_t^* \right\rangle e^{\ii  (s-t)\theta}, \nonumber \\
    &= \frac{1}{2 \pi} \frac{\Gamma(S+1)}{\Gamma(S+2)} \sum_{s,t=0}^{S} \delta_{s,t} = \frac{1}{2\pi},
\end{align}
where the value of $\left\langle V_s V_t^* \right\rangle$ was computed from \cref{eq:second_order_moment_V}, and where $\langle \cdot \rangle$ denotes an averaging over all vectors $\vvec{V}$ of the unit sphere. As expected, this averaged distribution is isotropic, and the random model thus satisfies the spherical symmetry condition expressed in \cref{eq:isotropic_average_pdf}.

\subsection{Average squared directionality in the random model}\label{appsubsec:directionality_random_model}
\par The average squared directionality in the random model can be computed from \cref{eq:random_model_correlation_function}. To do so, we use the Fourier series expansion of the Fejér kernel \cite{Stein2003_Fourier} which reads
\begin{equation}
    \frac{\sin^2\left[(S+1)\tau/2\right]}{(S+1)\sin^2(\tau/2)} = 1 + 2 \sum_{s=1}^{S} \left( 1 - \frac{s}{S+1} \right) \cos(s \tau).
\end{equation}
Inserting this expression into \cref{eq:directionality_squared_2} yields
\begin{align}\label{eq:random_model_w_intermediate_1}
    \langle \vvec{w}^2 \rangle &= \frac{2}{\pi} \int_0^\pi \dd\tau \,\cos(\tau) \Biggl\{ \frac{(S+1)}{(S+2)} \nonumber \\
    &+ \frac{1}{(S+2)}\sum_{s=1}^{S} \left( 1 - \frac{s}{S+1} \right) \cos(s \tau) \Biggr\}.
\end{align}
The integral of the first term cancels, while only the term $s=1$ in the sum contributes to the integral of the second term, because of the orthogonality relation of cosine functions. This leads to
\begin{align}\label{eq:random_model_w_final}
    \langle \vvec{w}^2 \rangle &= \frac{2}{\pi} \frac{S}{(S+1)(S+2)} \int_0^\pi \dd\tau \,\cos^2(\tau), \nonumber \\
    &= \frac{S}{(S+1)(S+2)},
\end{align}
which corresponds to \cref{eq:random_model_w} in the main text.

\bibliography{refs}

\end{document}